\documentclass[useAMS,usenatbib]{mn2e}
\usepackage{graphicx}    
\usepackage{dcolumn}     
\usepackage{bm}          
\usepackage{amssymb,amsmath}  
\usepackage{color}

\title[Measuring the transition to homogeneity with photometric redshift surveys]
      {Measuring the transition to homogeneity with photometric redshift surveys}
\author[D. Alonso et al.]
{D. Alonso$^1$, A. Bueno Belloso$^2$, F. J. S\'anchez$^3$, J. Garc\'{i}a-Bellido$^4$,
 E. S\'anchez$^3$\\
 $^1$Astrophysics, University of Oxford, DWB, Keble Road, Oxford, OX1 3RH, UK\\
 $^2$Institut f\"ur Theoretische Physik, Universit\"at Heidelberg, Philosophenweg 16
     69120 Heidelberg, Germany\\
 $^3$Centro de Invstigaciones Energ\'eticas, Medioambientales y Tecnol\'ogicas (CIEMAT), Madrid,
     Spain \\
 $^4$Instituto de F\'isica Te\'orica, Universidad Aut\'onoma de Madrid (IFT-UAM/CSIC), Madrid,
     Spain \\
}
\begin{document}

  \date{\today}
  \pagerange{1--14} \pubyear{2013}
  \maketitle
  \label{firstpage}

  \begin{abstract}
    We study the possibility of detecting the transition to homogeneity using photometric redshift
    catalogs. Our method is based on measuring the fractality of the projected galaxy distribution,
    using angular distances, and relies only on observable quantites. It thus provides a way to
    test the Cosmological Principle in a model-independent unbiased way. We have tested our method
    on different synthetic inhomogeneous catalogs, and shown that it is capable of discriminating
    some fractal models with relatively large fractal dimensions, in spite of the loss of
    information due to the radial projection. We have also studied the influence of the redshift
    bin width, photometric redshift errors, bias, non-linear clustering, and surveyed area, on
    the angular homogeneity index $H_2(\theta)$ in a $\Lambda$CDM cosmology. The level to which an
    upcoming galaxy survey will be able to constrain the transition to homogeneity will depend
    mainly on the total surveyed area and the compactness of the surveyed region. In particular, a
    Dark Energy Survey (DES)-like survey should be able to easily discriminate certain fractal models with fractal
    dimensions as large as $D_2=2.95$. We believe that this method will have relevant applications
    for upcoming large photometric redshift surveys, such as DES or the Large Synoptic Survey Telescope (LSST).
  \end{abstract}


  \section{Introduction}
  \label{sec:introduction}
    The standard model of cosmology is based on the so-called Cosmological Principle, which states
    that, on sufficiently large scales, the Universe must be homogeneous and isotropic (i.e.:
    statistical averages, such as the mean matter density, must be translationally and rotationally
    invariant). The validity of the Cosmological Principle is of paramount
    relevance for the standard model, and therefore it is extremely important to verify it using
    unbiased observational probes. In this model, the homogeneous regime is only reached
    asymptotically on large scales, and is evidently not realized on small scales, due to the form
    of the spectrum of matter perturbations and to their evolution via gravitational collapse.    
    The primordial spectrum of metric perturbations is predicted to be almost scale-invariant
    within the theory of inflation, and this result is supported by CMB data
    \citep{planck-collaboration:2013a}. The evolution of these perturbations after inflation varies
    the shape of the spectrum, but we can still expect a certain degree of inhomogeneity on all
    scales. In any case, the gradual transition to homogeneity is well understood and can be
    modelled and compared with observational data \citep{Bagla:2007tv}.

    Large-scale homogeneity is usually assumed without proof when analyzing certain cosmological
    probes \citep{Durrer:2011gq}. This is often a reasonable approach, since it would
    not be possible to obtain many observational constraints without doing so. However, in order to
    be able to rely on these constraints, we must verify the validity of the Cosmological Principle
    independently in an unbiased way. Along these lines, different groups have argued that the
    Universe might in fact not reach a homogeneous regime on large scales, and that instead it
    behaves like a fractal \citep{1992PhR...213..311C,1997cdc..conf...24P,1997EL.....39..103M,
    1998PhR...293...61S,1999ApJ...514L...5J,2009A&A...508...17S,2011EL.....9659001S}, while other
    groups claim the opposite result: the predictions of the standard $\Lambda$CDM model are in
    perfect agreement with the observational data, and homogeneity is indeed reached on scales of
    $\mathcal{O}(100)\,{\rm Mpc}/h$ \citep{Hogg:2004vw,1994ApJ...437..550M,1997NewA....2..517G,
    1997ApL&C..36...59S,1998MNRAS.298.1212M,1998A&A...334..404S,Amendola:1999gd,
    2000MNRAS.318L..51P,2001A&A...370..358K,2005MNRAS.364..601Y,2009MNRAS.399L.128S,
    Scrimgeour:2012wt,2013MNRAS.434..398N}.
    
    The disparity between these two results seems to stem from the differences in the analysis
    methods. On the one hand it is desirable to use methods that are, as far as possible, free of
    assumptions, especially regarding the property you want to measure. However, in the case of
    the validity of the Cosmological Principle, this is not an easy task, since homogeneity must
    sometimes be assumed in order to cope with certain observational effects. These issues will
    be further explained in section \ref{sec:fractality}. At the end of the day, we must ensure
    that the method used is able to distinguish homogeneous from non-homogeneous models to a
    reasonable level of precision. A robust and popular method to study the transition to
    homogeneity in the matter density field at late times is to analyze the fractality of the
    galaxy distribution in a redshift survey. Furthermore, fractal dimensions can be used
    to quantify clustering, since they depend on the scaling of the different moments of galaxy
    counts in spheres, which in turn are related to the $n$-point correlation functions.
    
    As has been said, the homogeneous regime is reached, within the standard $\Lambda$CDM model, at
    very large scales, and therefore a large survey volume is
    necessary in order to safely claim a detection of this transition. In this sense, photometric
    galaxy redshift surveys such as DES \citep{Abbott:2005bi} provide a unique
    oportunity for this study, since they are able to observe large numbers of objects distributed
    across wide areas and to further redshifts than their spectroscopic counterparts. The main
    caveat of these surveys is that, due to the limited precision in the redshift determination,
    much of the radial information is lost, and we are only able to study angular clustering in
    different thick redshift slices. Hence, in order to study the fractality of the galaxy
    distribution with a photometric survey, the methods and estimators used in previous analyses
    must be adapted to draw results from angular information alone. One advantage of this approach
    is that, since angular positions are pure observables (unlike three-dimensional distances,
    which can only be calculated assuming a fiducial cosmology), the results obtained are
    completely model independent. In this paper we propose an observable, the angular homogeneity
    index $H_2(\theta)$, which could be used by photometric surveys in the near-future to study the
    fractal structure of the galaxy distribution.
    
    In section \ref{sec:fractality} we describe one of the most popular observables used in the
    literature to study the fractality of the galaxy distribution, the correlation dimension $D_2$,
    and propose a way to adapt this quantity to the data available in a photometric galaxy survey.
    Here, the angular homogeneity index $H_2(\theta)$ is presented and modelled in the $\Lambda$CDM
    cosmology. In section \ref{sec:measuring_h} we analyze the fractality of a set of $\Lambda$CDM
    mock galaxy surveys using the method described before and study the different effects that may
    influence this measurement. The ability of our method to distinguish different inhomogeneous
    models is studied in section \ref{sec:robustness} by using it on different simulated
    inhomogeneous distributions. Finally the main results of this work are discussed in section
    \ref{sec:discussion}.

  \section{Fractality}
  \label{sec:fractality}
    There exist different statistical quantities that can be studied in order to quantify
    the fractality of a point distribution, such as the box-counting dimension, the different
    Minkowski-Bouligand dimensions or the lacunarity of the distribution (see 
    \citet{Martinez:2002} for a review of these). Of these, we will focus here on the
    Minkowski-Bouligand dimension of order 2, also called the correlation dimension, for the
    three-dimensional case. A simple modification of this observable will then allow us to
    study the fractality of the distribution from its angular projection.

    \subsection{The fractal dimension}
    \label{ssec:fractal_dimension}
      For a given point distribution, let us define the correlation integral $C_2(r)$ as the
      average number of points contained by spheres of radius $r$ centerered on other points of
      the distribution. For an infinite random point process in three dimensions, this quantity
      should grow like the volume
      \begin{equation}
        C_2(r)\propto r^3,
      \end{equation}
      and thus we define the correlation dimension of the point process as
      \begin{equation}\label{eq:d2}
        D_2(r)\equiv \frac{d\log C_2}{d\log r}.
      \end{equation}
      Hence, if the galaxy distribution approaches homogeneity on large scales, $D_2$ must tend
      to 3 for large $r$.

      For the canonical $\Lambda$CDM model, departures from this value are due to two different
      reasons. First, since the galaxy distribution is clustered due to the nature of gravitational
      collapse, there exists an excess probability of finding other galaxies around those used as
      centers to calculate $D_2$. Secondly, in practice, the point distributions under study are
      finite in size, and this introduces an extra contribution due to shot-noise. These two
      contributions have been modelled by \citet{Bagla:2007tv} for the correlation integral:
      \begin{align}
        & C_2(r)=\overline{N}(r)+[\Delta C_2(r)]_{\rm cluster}+[\Delta C_2(r)]_{\rm sn}\\\nonumber
        & [\Delta C_2(r)]_{\rm cluster} = \overline{N}(r)\,\bar{\xi}(r),\\\nonumber
        & [\Delta C_2(r)]_{\rm sn} = 1,
      \end{align}
      and the correlation dimension
      \begin{align}
        & D_2(r) = 3+[\Delta D_2(r)]_{\rm cluster}+[\Delta D_2(r)]_{\rm sn}\\\nonumber
        & [\Delta D_2(r)]_{\rm cluster}=-3\,\frac{\bar{\xi}(r)-\xi(r)}{1+\bar{\xi}(r)},\\\nonumber
        & [\Delta D_2(r)]_{\rm sn} = -\frac{3}{\overline{N}(r)},
      \end{align}
      where $\bar{\xi}(r)$ is the volume-averaged two-point correlation function of the
      distribution
      \begin{equation}
        \bar{\xi}(r)\equiv\frac{3}{r^3}\int_0^r\,s^2\xi(s)\,ds
      \end{equation}
      and $\overline{N}(r)\equiv 4\pi\bar{n}\,r^3/3$ is the average number of objects inside
      spheres of radius $r$. Since the contribution due to shot noise will always be present in    
      any finite distribution, we will substract it by hand in this work, and focus only on the
      clustering term.

    \subsection{The angular homogeneity index}
    \label{ssec:homogeneity_index}
      The observables described in the previous section can be adapted straightforwardly to point
      distributions projected onto the 2-dimensional sphere. Instead of spheres of radius $r$, we
      will consider here spherical caps of radius $\theta$.

      In analogy with the three-dimensional case, we can define the angular correlation integral
      $G_2(\theta)$ as the average number of points inside spherical caps of radius $\theta$
      centered on other points of the distribution. For a homogeneous distribution, this quantity
      should grow like the ``volume'' (i.e. the solid angle) inside these spherical caps. However,
      since this volume $V(\theta)=2\pi\,(1-\cos\theta)$ does not grow as a simple power of
      $\theta$, a logarithmic derivative of $G_2$ with respect to $\theta$ would not capture the
      approach to homogeneity in a simple manner, independent of the angular radius. Therefore
      we have preferred to define the homogeneity index $H_2(\theta)$ as the logaritmic derivative
      with respect to the volume:
      \begin{equation}
        H_2(\theta)=\frac{d\log G_2(\theta)}{d\log V(\theta)},
      \end{equation}
      which should tend to 1 if homogenity is reached.

      As in the three-dimensional case, these quantities can be modelled for a finite weakly
      clustered distribution:
      \begin{align}
        & G_2(\theta)=1+\overline{N}(\theta)\,[1+\bar{w}(\theta)],\\
        & H_2(\theta)=1-\frac{\bar{w}(\theta)-w(\theta)}{1+\bar{w}(\theta)}-
                    \frac{1}{\overline{N}(\theta)},\\
        & \overline{N}(\theta)\equiv2\pi\,\bar{\sigma}\,(1-\cos\theta),
      \end{align}
      where $\bar{\sigma}$ is the mean angular number density of the distribution, $w(\theta)$ is
      the angular two-point correlation function and $\bar{w}(\theta)$ is defined in analogy to
      $\bar{\xi}(r)$:
      \begin{equation}
        \bar{w}(\theta)\equiv\frac{1}{1-\cos\theta}\int_0^{\theta}w(\theta')\,
                             \sin\theta'\,d\theta'.
      \end{equation}
      In this paper we will be interested in the departure of $H_2$ from its homogeneous value:
      $\Delta H_2(\theta) \equiv 1-H_2(\theta)$. This quantity must not be mistaken with the
      statistical error on the determination of $H_2$, which we label $\sigma_{H_2}$ here.
      
    \subsection{Measuring the transition to homogeneity}\label{ssec:measure}
      When trying to measure the fractal dimension or the homogeneity index from a realistic galaxy
      survey, different complications arise, mainly related with the artificial observational
      effects induced on the galaxy distribution, which must be correctly disentangled from the
      clustering pattern and from a possible fractal-like structure. For instance, unless a
      volume-limited sample is used, we will have to deal with a non-homogeneous radial
      selection function. Furthermore, the angular distribution of the survey galaxies will
      always contain imperfections, which may come, for example, from survey completeness,
      fiber collisions and star contamination for a spectroscopic survey, or CCD saturation
      in photometric catalogs. Although it would be desirable to be able to deal with these
      effects without making any extra assumptions about the true galaxy distribution, in
      order to make sure that our method of analysis is not biased towards a homogeneous
      solution, this is often not possible. The most popular method to circumvent these issues
      in the calculation of the two-point correlation function, is to use random
      catalogs that incorporate the same artificial effects as the data, and a similar approach
      may be used for our purposes. In this work we have considered three different estimators
      for $D_2$, which are described below.

      For the $i$-th galaxy of the survey, let us define $n_i^d(<r)$ as the number of galaxies in
      the survey inside a sphere of radius $r$ centered around $i$, and $n_i^r(<r)$ as the same
      quantity for an unclustered random distribution. For $N_c$ galaxies used as sphere centres,
      we can define the scaled counts-in-spheres $\mathcal{N}(r)$ as
      \begin{equation}
        \mathcal{N}(r)\equiv\frac{1}{N_c}\sum_{i=1}^{N_c}\frac{n_i^d(<r)}{f_r\,n_i^r(<r)},
      \end{equation}
      where $f_r\equiv D/R$ is the ratio of the number of galaxies in the survey to the number of
      points in the random catalog. Varying the prescription to estimate $n_i^r$ and to select
      galaxies as sphere centres, we can define three different estimators:
      \begin{enumerate}
        \item {\bf E1}. In the most conservative case, in order to avoid any assumptions about the
              galaxy distribution, around each galaxy we may only use spheres that fit fully inside
              the surveyed volume. Thus, the number of centers will be a function of $r$. Also,
              assuming that there are no other artificial effects in the galaxy distribution, we
              may estimate $n_i^r(<r)$ theoretically as
              \begin{equation}
                n_i^r(<r)=\overline{N}(r)=\frac{4\pi}{3}r^3\,\bar{n}_d,
              \end{equation}
              where $\bar{n}_d$ is the survey's mean number density and we have assumed $f_r=1$.
        
              This estimator is very idealistic and problematic to use in a realistic scenario, in
              which observational effects are not negligible.
        \item {\bf  E2}. While still using only complete spheres, we may use a random catalog that
              incorporates the same observational effects as the data to estimate $n_i^r(<r)$. This
              way we are able to study the fractality of a survey that is not volume limited, as
              well as to incorporate small-scale observational effects, without assuming anything
              about the galaxy distribution outside the survey.
        \item {\bf E3}. In order to maximize the use of the survey data, we may use all galaxies as
              sphere centres for all radii. This implies using spheres that lie partly outside the
              surveyed region, a fact that is accounted for by using a random catalog to estimate
              $n_i^r(<r)$ in those same spheres.
      \end{enumerate}
      Once $\mathcal{N}(<r)$ is estimated, it can be directly related to the correlation integral
      through
      \begin{equation}
        C_2(r)=\overline{N}(r)\,\mathcal{N}(<r)-1,
      \end{equation}
      where we have explicitly substracted the shot-noise contribution. $C_2$ can then be used
      to calculate the fractal dimension through equation (\ref{eq:d2}).

      Two final points must be made regarding the use of random catalogs in order to deal with
      observational effects. First, we must be very careful to incorporate in these only purely
      artificial effects in order to minimize a possible bias of our estimator towards homogeneity.
      Even doing so, it is clear that the only way to avoid this bias is by using {\bf E1} on a
      volume-limited survey using only regions that are 100\% complete and free of any
      observational issues, however this is too restrictive for any realistic galaxy survey. This
      approach is impractical and, therefore, we have only considered the estimators {\bf E2} and
      {\bf E3} in the rest of this work. These estimators contain an extra
      contribution due to the finiteness of the random catalogs used to estimate $n_i^r(<r)$ (i.e.,
      they are biased). This bias can only be suppressed by using many times more random objects
      than points in the data ($f_r\ll1$). Note that in the limit of infinite random objects,
      and in the absence of artificial inhomogeneities, {\bf E1} and {\bf E2} are equivalent.      
      
      As is shown in section \ref{sec:robustness}, we have tested that the use of the least
      conservative estimator {\bf E3} does not introduce any significant bias towards homogeneity
      by using it on explicitly inhomogeneous data. Since this estimator makes the most efficient
      use of the data, we have used it for most of the analysis presented in sections
      \ref{sec:measuring_h} and \ref{sec:robustness}, and it will be assumed unless otherwise
      stated.
      
      The estimators for $H_2(\theta)$ from a finite projected distribution can be constructed in
      analogy with the ones presented above for three-dimensional distributions. In this
      case, they are based on calculating the scaled counts-in-caps
      \begin{equation}
        \mathcal{N}(<\theta)\equiv\frac{1}{N_c}\sum_{i=1}^{N_c}
                                  \frac{n_i^d(<\theta)}{f_r\,n_i^r(<\theta)},
      \end{equation}
      using different prescriptions for $N_c$ and $n_i^r(<\theta)$.
      
    \subsection{Modelling $H_2(\theta)$}
    \label{ssec:modelling_htheta}
      As we have seen, the angular homogeneity index is directly related, to first order, with the
      angular two-point correlation function $w(\theta)$. Thus, in order to forecast the ability of
      a given galaxy survey to study the transition to homogeneity, we need to be able to model
      $w(\theta)$ correctly. This is extensively covered in the literature (e.g.
      \citet{Crocce:2010qi}), therefore we will only quote the main results here. The angular
      correlation function is related to the anisotropic 3D correlation function $\xi(r,\mu)$
      through
      \begin{equation}
        w(\theta)=\int dz_1\,\phi(z_1)\int dz_2\,\phi(z_2)\,\xi(r,\mu),
      \end{equation}
      where $\phi(z)$ is the survey selection function and
      \begin{align}
        & r=\left[\chi^2(z_1)+\chi^2(z_2)-2\,\chi(z_1)\chi(z_2)\cos\theta\right]^{1/2}, \nonumber\\
        & \mu=\frac{|\chi^2(z_1)-\chi^2(z_2)|}{[(\chi^2(z_1)+\chi^2(z_2))^2-4\chi^2(z_1)\chi^2(z_2)
              \cos^2\theta]^{1/2}}.
      \end{align}
      Here $\chi(z)$ is the radial comoving distance to redshift $z$ given by
      \begin{equation}
        \chi(z)\equiv\int_0^z\frac{dz'}{H(z')}.
      \end{equation}

      The selection function in these equations must be normalized to unity
      \begin{equation}
        \int_0^{\infty}\phi(z)dz=1,
      \end{equation}
      and the effects of a non-zero photometric redshift error can be included in the selection
      function by convolving the true-$z$ $\phi(z)$ with the photo-$z$ probability distribution
      function. In the ideal case of Gaussianly distributed redshift errors, and for a redshift    
      bin $z_0 < z < z_f$, this is \citep{Asorey:2012rd}
      \begin{equation}
        \phi_{\rm photo}(z)=\phi_{\rm true}(z)\left(
                            {\rm erf}\left[\frac{z_f-z}{\sqrt{2}\sigma_z}\right]-
                            {\rm erf}\left[\frac{z_0-z}{\sqrt{2}\sigma_z}\right]
                            \right),
      \end{equation}
      where $\sigma_z$ is the rms Gaussian photo-$z$ error.

        \begin{figure*}
          \centering
          \includegraphics[width=0.45\textwidth]{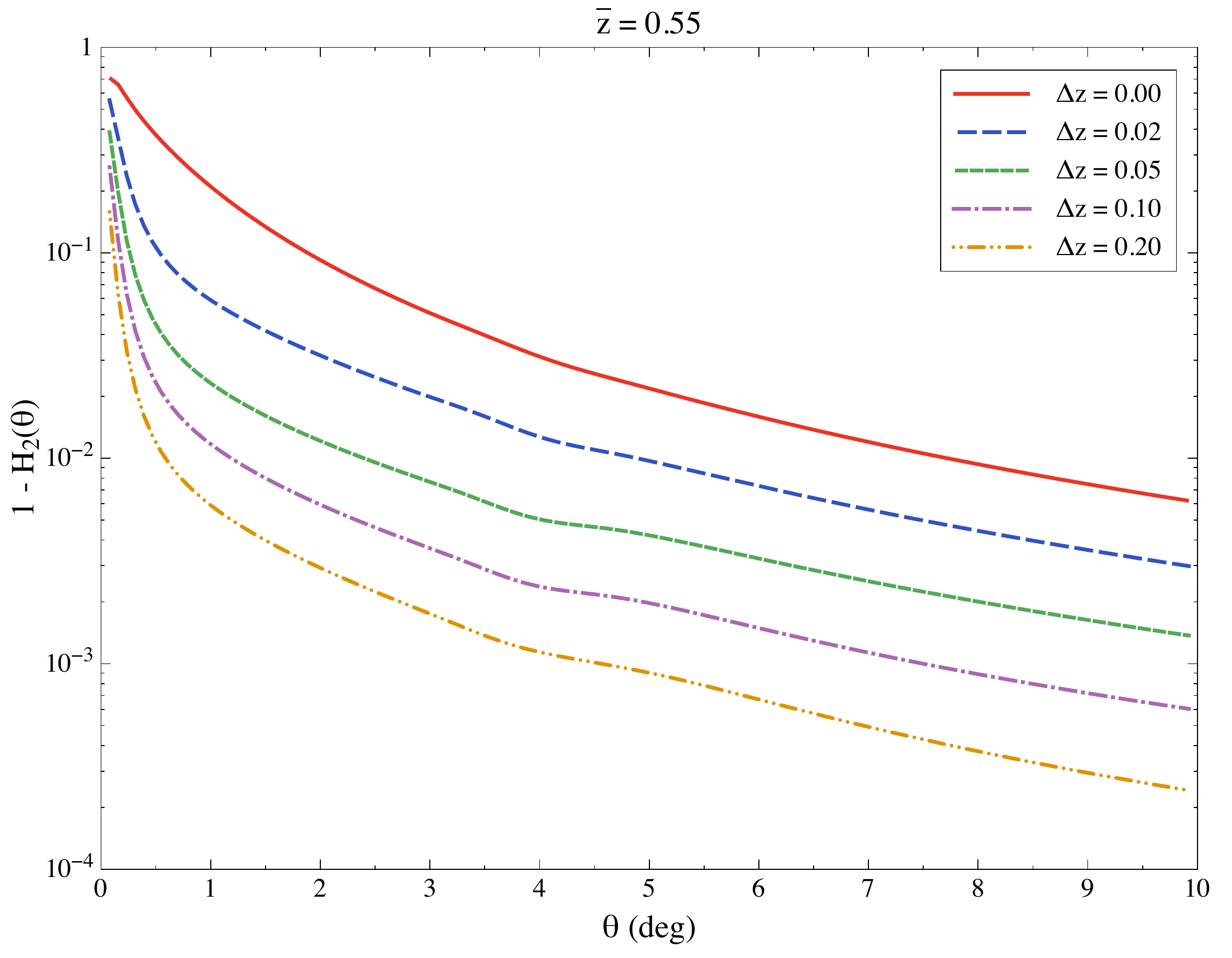}
          \includegraphics[width=0.45\textwidth]{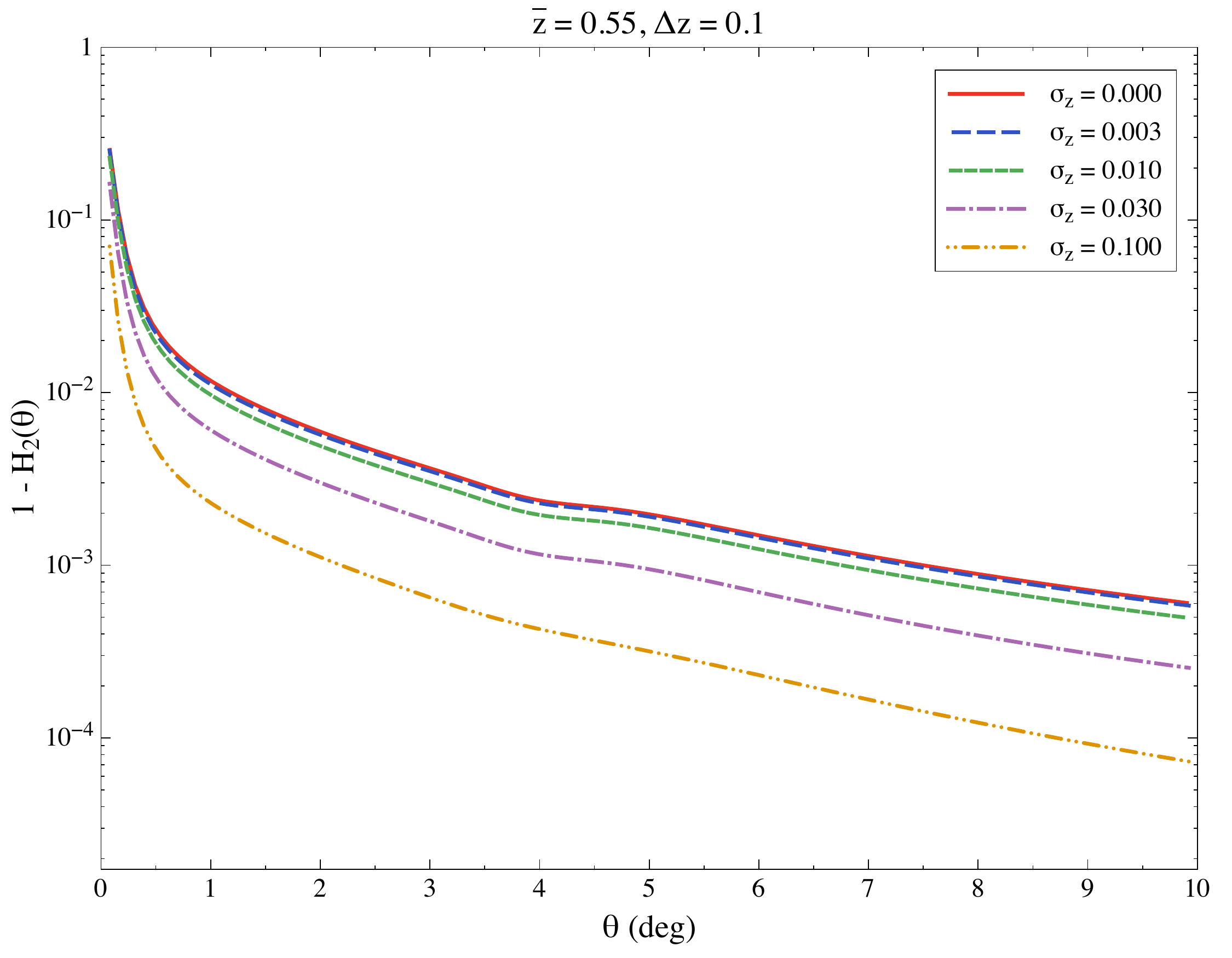}
          \caption{$\Delta H_2(\theta)$ as a function of $\theta$ for varying redshift bin size
                   (left panel) and photometric redshift uncertainty (right panel). The use of
                   thick redshift bins and photometric redshifts produces a more homogeneous
                   distribution when projected on the sphere, reducing the amplitude of the
                   correlation.}
          \label{fig:project}
        \end{figure*}
      The three-dimensional correlation function is related to the power spectrum through a Fourier
      transform
      \begin{equation}
        \xi({\bf r})=\frac{1}{(2\pi)^3}\int\,dk^3\,e^{i\,{\bf k}\cdot{\bf r}}P({\bf k}).
      \end{equation}
      The following model was used for the redshift-space power spectrum
      \begin{equation}
        P_s(k,\mu_k,z) = b^2(z)\,(1+\beta(z)\mu_k^2)^2\,P_r(k,z),
      \end{equation}
      where $P_r(k,z)$ is the real-space power spectrum, $b(z)$ is the linear galaxy bias,
      $\beta=f/b$ is the redshift-distortion parameter and $\mu_k$ is the cosine of the angle
      between the wave vector ${\bf k}$ and the line of sight. Non-linearities were taken into
      account by using the HALOFIT prediction \citep{2003MNRAS.341.1311S}. The power spectra
      used for the theoretical predictions were provided by {\tt CAMB} \citep{2000ApJ...538..473L}.
      For the figures shown in this section we used the flat $\Lambda$CDM parameters
      \begin{equation}\label{eq:cosmopar}
        (\Omega_M,\Omega_{\Lambda},\Omega_b,h,\sigma_8,n_s)=(0.3,0.7,0.049,0.67,0.8,0.96)
      \end{equation}
      as a fiducial cosmology.

      \subsubsection{Projection effects and bias}
      \label{sssec:projection_and_bias}

        Different effects have an influence in the way the galaxy distribution approaches
        homogeneity. In the case that concerns us, that of data projected on the sphere, this
        projection effectively homogenizes the distribution. This is easy to understand: consider
        a pair of galaxies subtending a small angle but separated by a large radial distance.
        While they are far away, and therefore almost uncorrelated, they appear close when
        projected. This effect is obviously larger for wider redshift bins, and therefore
        $H_2(\theta)$ will approach 1 on smaller scales as we increase the binwidth. This effect
        is shown in the left panel of figure \ref{fig:project}. The effect of a large photo-$z$
        error is similar: the photo-$z$ shifts galaxies from adjacent redshift bins, effectively
        making the bin width larger (see right panel of figure \ref{fig:project}).
        
\begin{figure}
          \centering
          \includegraphics[width=0.45\textwidth]{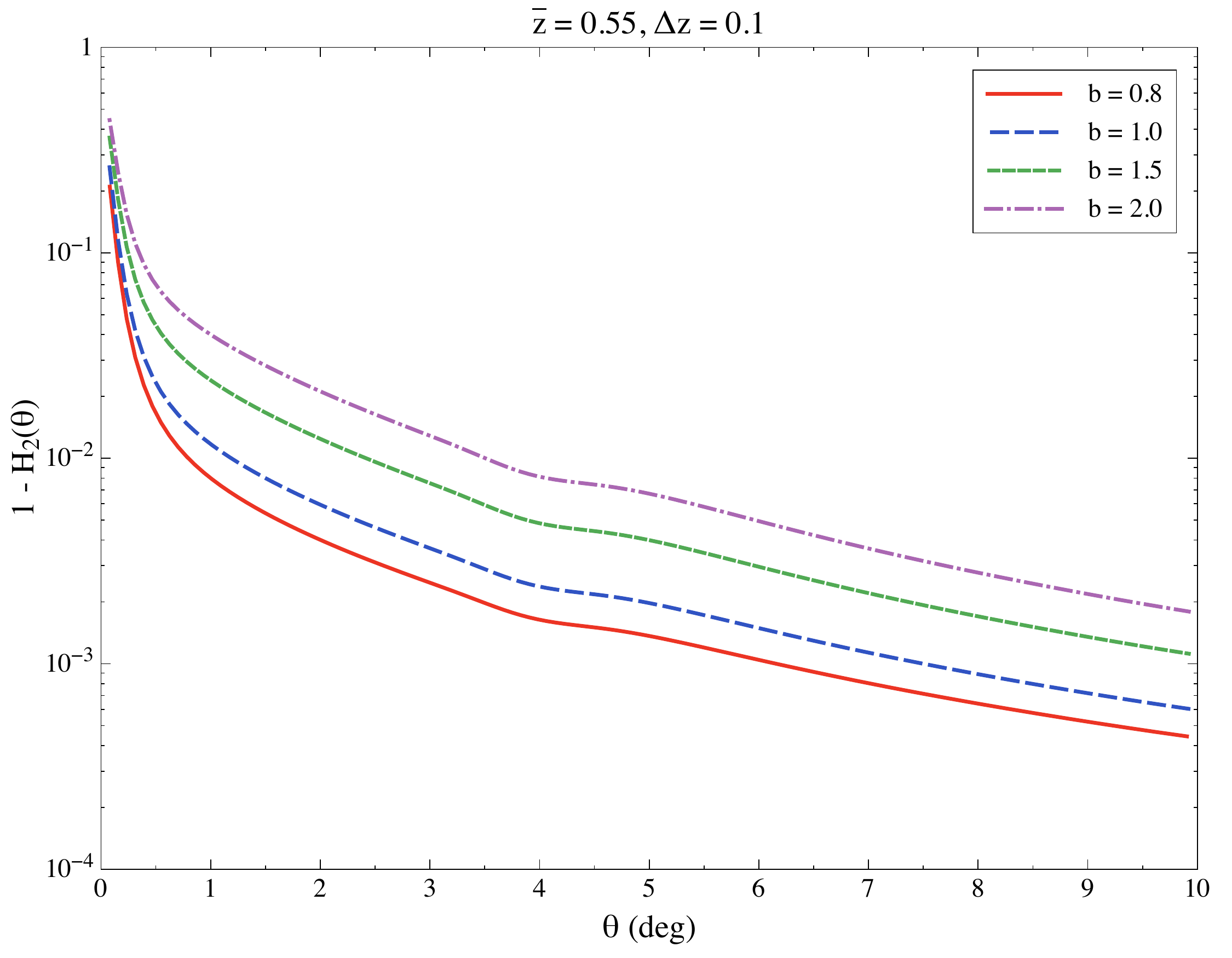}
          \caption{$\Delta H_2(\theta)$ as a function of $\theta$ for varying galaxy bias. A biased
                   galaxy population will be more tightly clustered, and therefore will
                   show a more evident departure from homogeneity.}
          \label{fig:bias}
        \end{figure}

        On the other hand, galaxy bias modifies the homogeneity index in the opposite way. A
        positively biased population ($b>1$) is more strongly clustered and therefore will
        reach the homogeneous regime on larger scales. This can be seen in figure \ref{fig:bias}.

      \subsubsection{Non-linearities}
      \label{sssec:non_linearities}
        \begin{figure}
          \centering
          \includegraphics[width=0.45\textwidth]{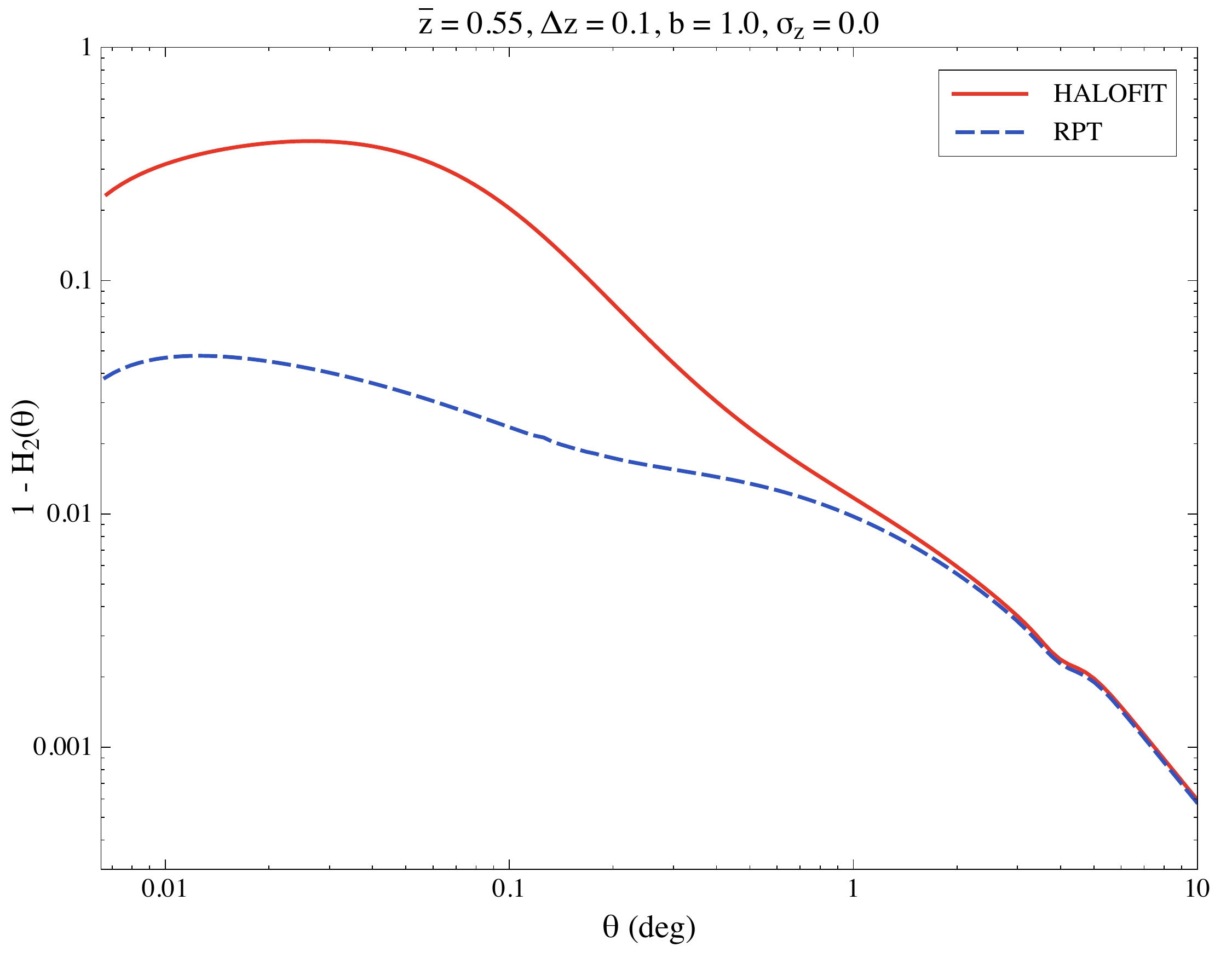}
          \caption{$\Delta H_2(\theta)$ as a function of $\theta$ for two different non-linearities
                   prescriptions. Since the $H_2(\theta)$ depends on an integral quantity, it
                   contains information about small scales. Thus, it is important to describe
                   non-linear effects correctly. The solid red line corresponds to the
                   prediction using the HALOFIT fitting formula, which fits well our
                   mock catalogs. The prediction including only the damping of the BAO
                   wiggles (dashed blue line) overpredicts $H_2(\theta)$ on small
                   scales, although this offset decreases for larger angles.}
          \label{fig:nlin}
        \end{figure}
        As we said, homogeneity is reached in the standard cosmological model on relatively large
        scales. Therefore one might think that the modelling of the small-scale non-linear effects
        should be irrelevant. However, the angular homogeneity index (or the correlation dimension
        in 3D) depends on an integral quantity (number counts \emph{inside} spheres), and therefore
        contains information about those small scales which may propagate to larger angles.
        
        This is shown in figure \ref{fig:nlin}, where the angular homogeneity index for the bin
        $z\in(0.5,0.6)$ has been plotted using different prescriptions to describe non-linearites.
        The solid red line shows the prediction using the HALOFIT fitting formula (which
        provides the best fit to the mock data in section \ref{sec:measuring_h}). The dashed blue
        line corresponds to the prediction in renormalized perturbation theory (RPT)
        \citep{Crocce:2006mi} of the damping of the BAO wiggles due to non-linear motions, given by
        \begin{equation}
          \Delta P_{\rm wiggles}^{NL}(k)=\Delta P_{\rm wiggles}^{L}(k)\,\exp(-\sigma_v^2k^2/2),
        \end{equation}
        where $\Delta P_{\rm wiggles}$ is the BAO contribution to the power spectrum and
        \begin{equation}
          \sigma_v = \frac{1}{6\pi^2}\int_0^{\infty}P^{L}(k)\,dk.
        \end{equation}        
        As can be seen, the extra clustering amplitude on small scales contributes as a visible
        offset in $H_2$ up to scales of $\mathcal{O}(1\,{\rm deg}\sim20\,{\rm Mpc}/h)$.

      \subsubsection{Dependence on cosmological parameters}
      \label{sssec:cosmological_parameters}
        \begin{figure*}
          \centering
          \includegraphics[width=0.45\textwidth]{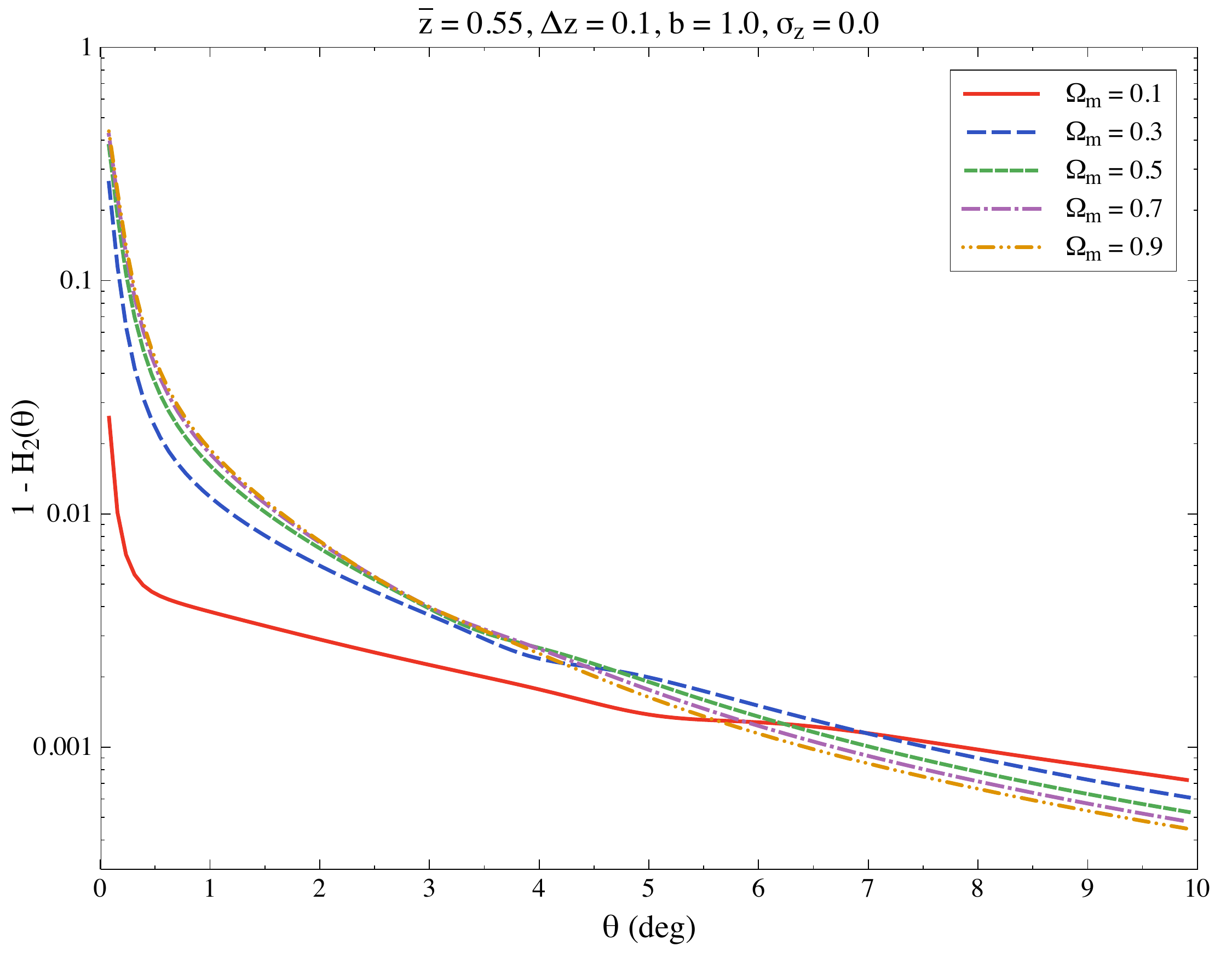}
          \includegraphics[width=0.45\textwidth]{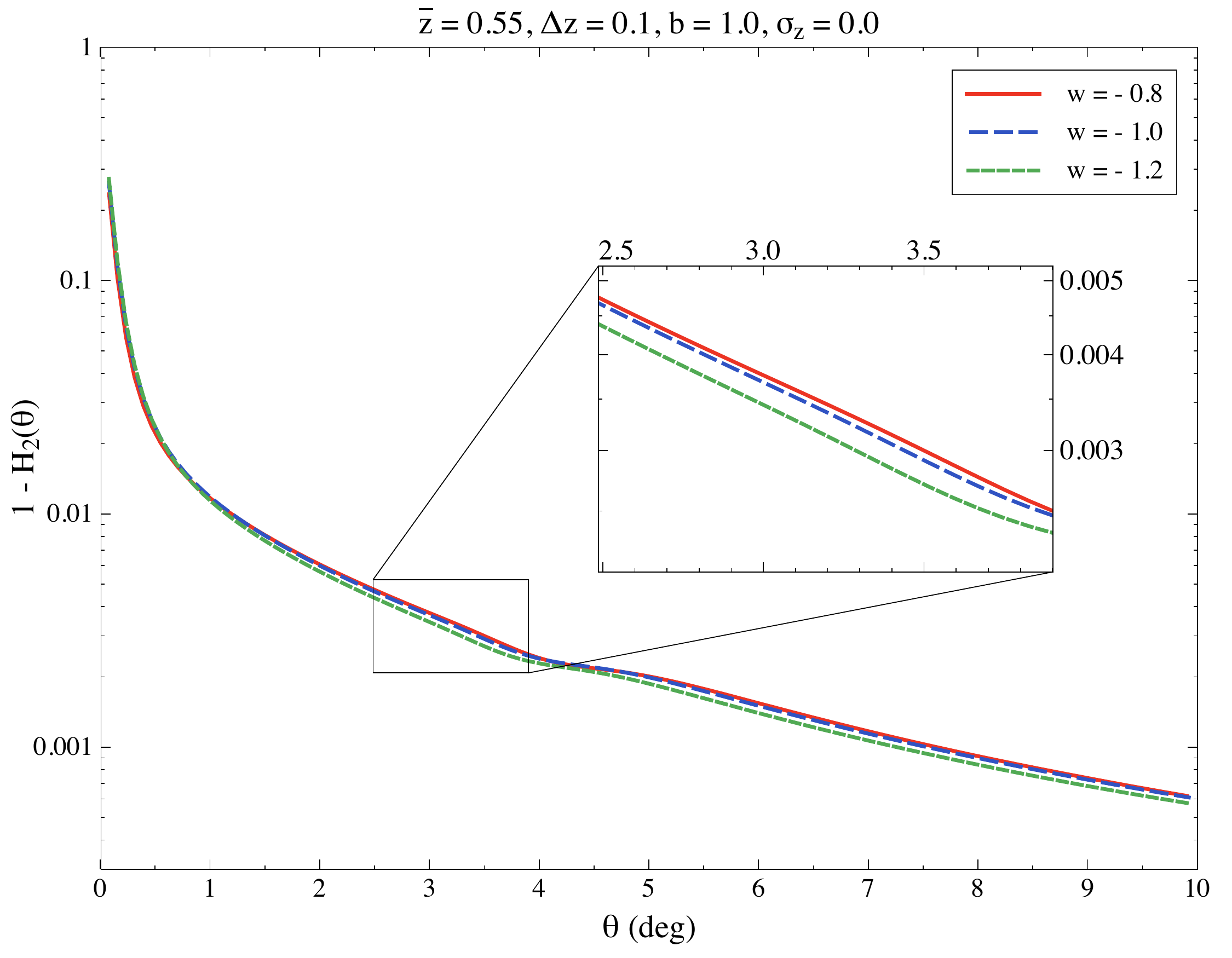}
          \caption{$\Delta H_2(\theta)$ as a function of $\theta$ for varying $\Omega_m$
                   (left panel) and dark energy equation of state $w$ (right panel).}
          \label{fig:cospar}
        \end{figure*}
        Since the evolution of the matter perturbations depends on the background cosmological
        parameters, we can expect that some cosmological models will approach homogeneity faster
        than others. We have studied this dependence using our model for $H_2(\theta)$. Our aim
        is not to use the form of $H_2(\theta)$ to obtain precise cosmological constraints,
        since we do not think that this quantity contains more information than the two-point
        correlation function $w(\theta)$, for which there exist many different methods in the
        literature \citep{Nock:2010gr,Sanchez:2010zg,deSimoni:2013oqa}. However, we think that
        a qualitative characterization of the homogeneity index for different types of models
        is instructive and may give us some hints about how model-independent our results are.
        
        In Fig.~\ref{fig:cospar} we have plotted $H_2(\theta)$ varying the values of the matter
        parameter $\Omega_M$ (left panel) and the dark energy equation of state $w$ (right panel),
        from their fiducial values (eq. \ref{eq:cosmopar}). As expected, larger values of
        $\Omega_M$ enhance the amplitude of inhomogeneities on small scales (i.e., make
        $\Delta H_2$ larger). Likewise, more negative values of $w$ accelerate the expansion
        and damp the growth of perturbations, shifting $H_2$ closer to~1. In any case we observe
        a mild dependence of $H_2$ on the cosmological parameters, and hence we expect that the
        results presented here should not vary qualitatively for any viable homogeneous
        cosmological model.
        
    \subsection{Defining homogeneity}
    \label{ssec:define_homogeneity}
      As has been discussed, even though the stantard cosmological model postulates a homogeneous  
      and isotropic Universe, this homogeneous regime is only approached asymptotically on
      large scales or in early times. Thus, there is no straightforward prescription to define the
      scale at which homogeneity is reached. Two different definitions have been used in the
      literature:
      \begin{itemize}
        \item One possibility is to define the scale of homogeneity as the scale at which the
              difference between our measurement of $D_2$ or $H_2$ (or, in general, any observable
              characterizing fractality) and its homogeneous value ($D_2=3$, $H_2=1$) is comparable
              with the uncertainty in this measurement. The caveat of this definition
              is that this uncertainty will depend on the characteristics of the survey (area,
              depth, number density, etc.), and therefore different surveys will measure a 
              different scale of homogeneity. However, this is possibly the most mathematically
              meaningful definition.
        \item Another approach is to define that homogeneity is reached when the measured fractal
              dimension is within a given arbitrary fraction of its homogeneous value. For example,
              \citet{Scrimgeour:2012wt} use a value of 1\%. The advantage of this definition is 
              that all surveys should measure the same scale of homogeneity, while its caveat is
              the arbitrariness of the mentioned fraction. Furthermore, using this kind of
              prescription would not be viable in our case, since, as we have seen, projection
              effects reduce the departure from homogeneity, and therefore the same fixed 
              fraction cannot be used for different bins.
      \end{itemize}
      For the data analyzed in the next section, we have chosen to follow the first prescription,
      defining the homogeneity scale $\theta_H$ as the angle at which
      \begin{equation}\label{eq:def_homog}
        \Delta H_2(\theta)\leq q\,\sigma_{H_2}(\theta),  
      \end{equation}
      where $\sigma_{H_2}$ is the error on $H_2$ and $q$ is an $\mathcal{O}(1)$ number. For this
      analysis we have used $q=1.96$ (i.e., assuming Gaussian errors, $\theta_H$ is the angle
      at which the measured $H(\theta)$ is away from 1 at 95\% C.L.). Note that the value of
      $\theta_H$ given by this definition should be interpreted as a lower bound on the
      scale of homogeneity, and not as a scale beyond which all inhomogeneities disappear.
      
      At the end of the day, the scale at which homogeneity is reached is not a well defined 
      quantity, nor is it of vital importance. Instead, the main aim of this kind of studies is to
      establish whether homogeneity is reached or not, focusing on defining the limits of our
      ability to detect a departure from large-scale homogeneity.

  \section{Measuring the homogeneity index}
  \label{sec:measuring_h}
    In order to assess the performance of the different estimators for $H_2(\theta)$ in a
    realistic scenario, we have used them on a set of simulated galaxy surveys corresponding
    to a canonical $\Lambda$CDM model.

    \subsection{Lognormal mock catalogs}
    \label{ssec:ln_mocks}
      \begin{figure}
        \centering
        \includegraphics[width=0.45\textwidth]{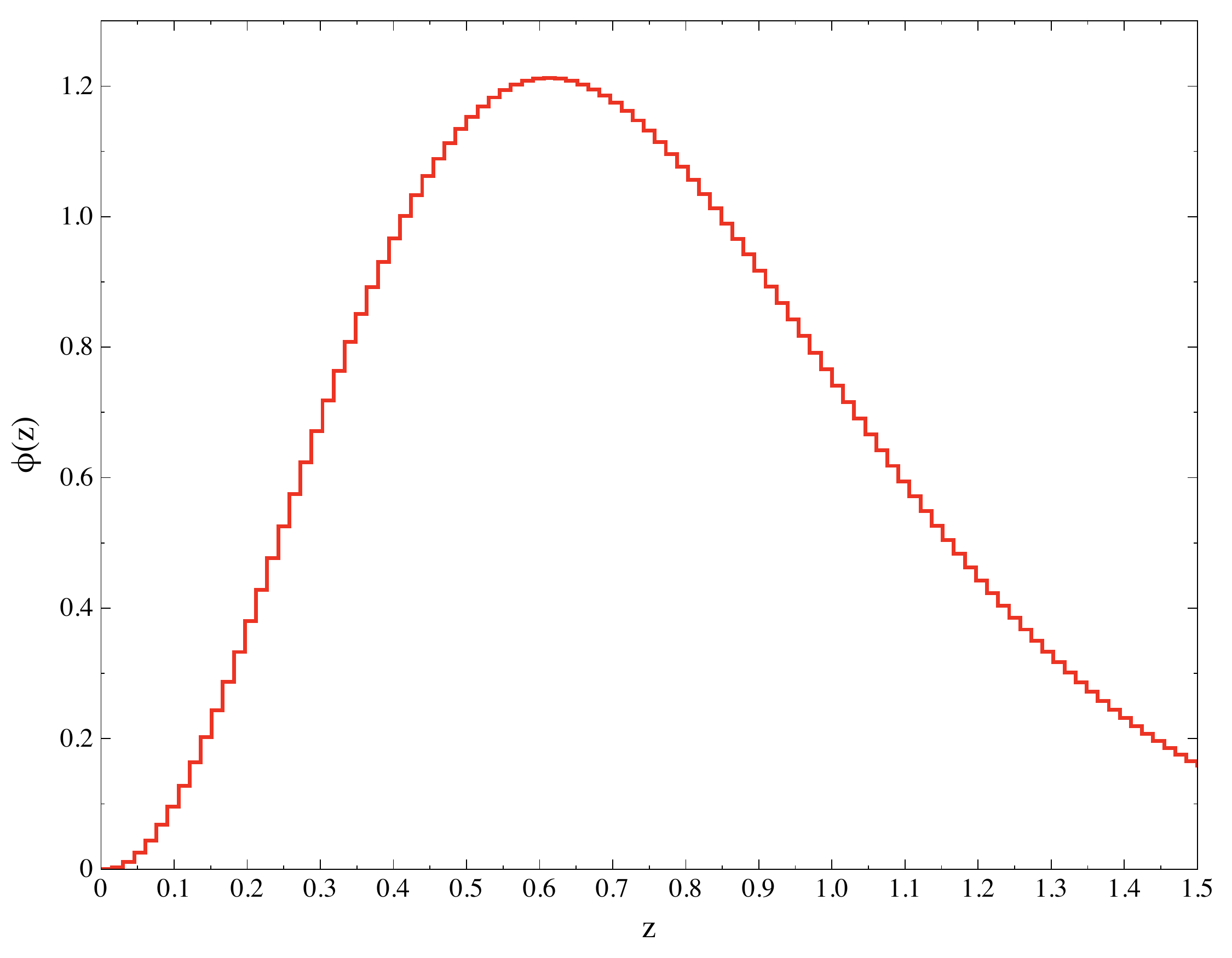}
        \caption{Selection function as a function of redshift used for the lognormal catalogs,
                 given by eq. (\ref{eq:phiz}).}
        \label{fig:phiz}
      \end{figure}

      Lognormal fields were proposed by \citep{Coles:1991if} as a possible way to describe the
      distribution of matter in the Universe. More interestingly for our purposes, lognormal
      fields provide an easy and fast method to generate realizations of the density field in
      order to produce large numbers of mock catalogs. This technique has been used by different
      collaborations to estimate statistical uncertainties and study different systematic effects  
      in galaxy surveys, and has been proven to be a remarkably useful tool. The physics and
      mathematics of lognormal realizations, as well as their limitations, have been widely
      covered in the literature \citep{Coles:1991if,White:2013psd}, and the specific method
      used to generate the catalogs used in this work is described in appendix
      \ref{sec:lognormals}.

      100 lognormal realizations were generated for the cosmological model of equation
      (\ref{eq:cosmopar}) inspired by the latest measurements by the Planck collaboration
      \citep{planck-collaboration:2013a}. Each catalog contains $1.2\times10^8$ galaxies
      distributed over one octant of the sky ($\simeq 5000\,{\rm deg}^2$) in the redshift range
      $0 < z < 1.4$ with a selection function
      \begin{equation}\label{eq:phiz}
        \phi_{\rm true}(z)\propto z^2\,e^{-\left(\frac{z}{0.5}\right)^{1.5}},
      \end{equation}
      shown in figure \ref{fig:phiz}. The density field was generated in a box of size
      $L_{\rm box}=3000\,{\rm Mpc}/h$ with a grid of size $N_{\rm side}=2048$, yielding a spatial
      resolution of $l_{\rm grid}\simeq1.5\,{\rm Mpc}/h$.
      All the catalogs contain redshift-space distortions, as described in appendix
      \ref{sec:lognormals}, and a Gaussian photometric redshift error was generated for each
      galaxy with $\sigma_z=0.03\,(1+z)$. Since the effect of a linear galaxy bias factor is well
      understood and very easy to model in theory, all the catalogs were generated with $b=1$.

    \subsection{Results}
    \label{ssec:results}
      \subsubsection{The angular homogeneity scale $\theta_H(z)$}
        \begin{figure*}
          \centering
          \includegraphics[width=0.3\textwidth]{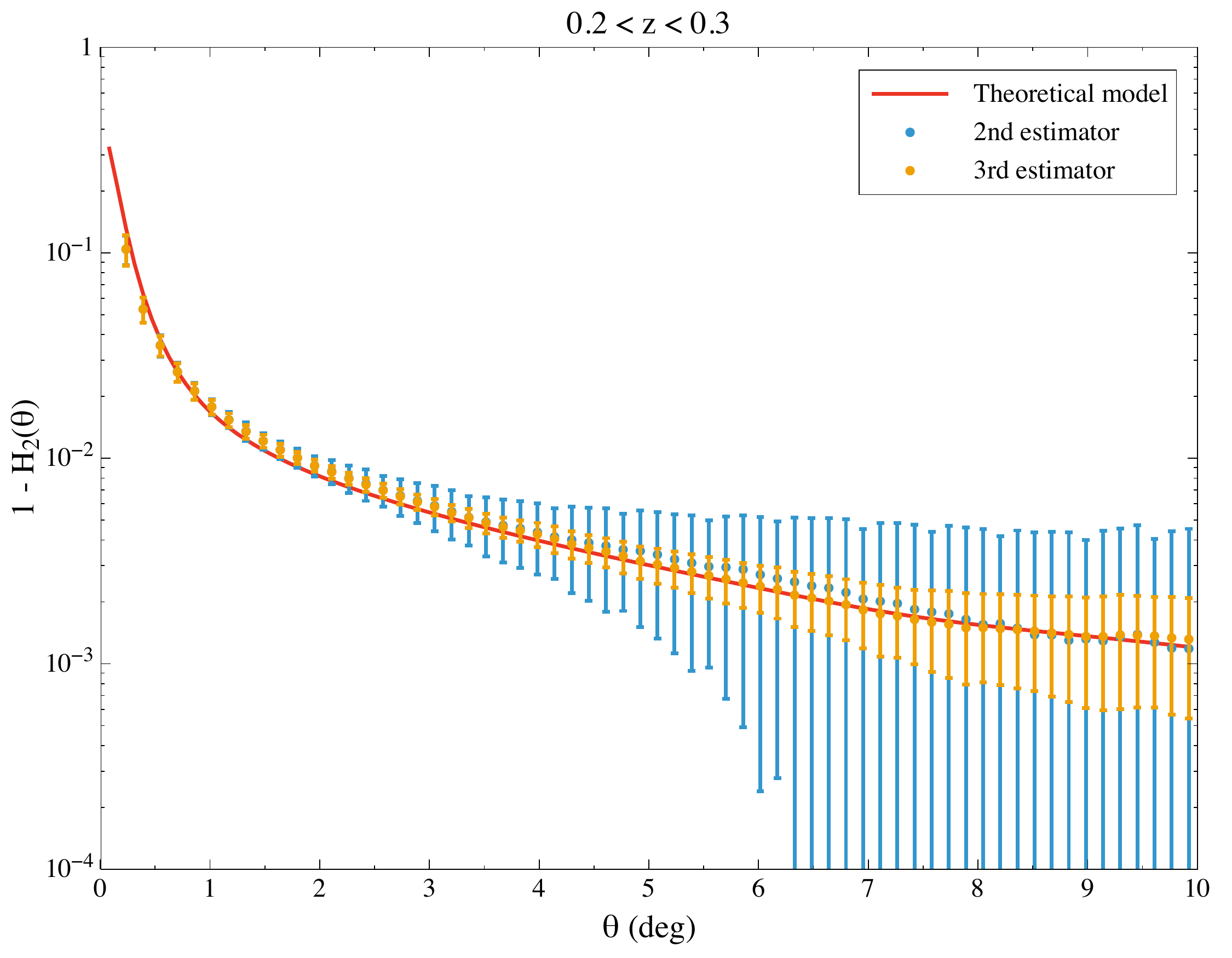}
          \includegraphics[width=0.3\textwidth]{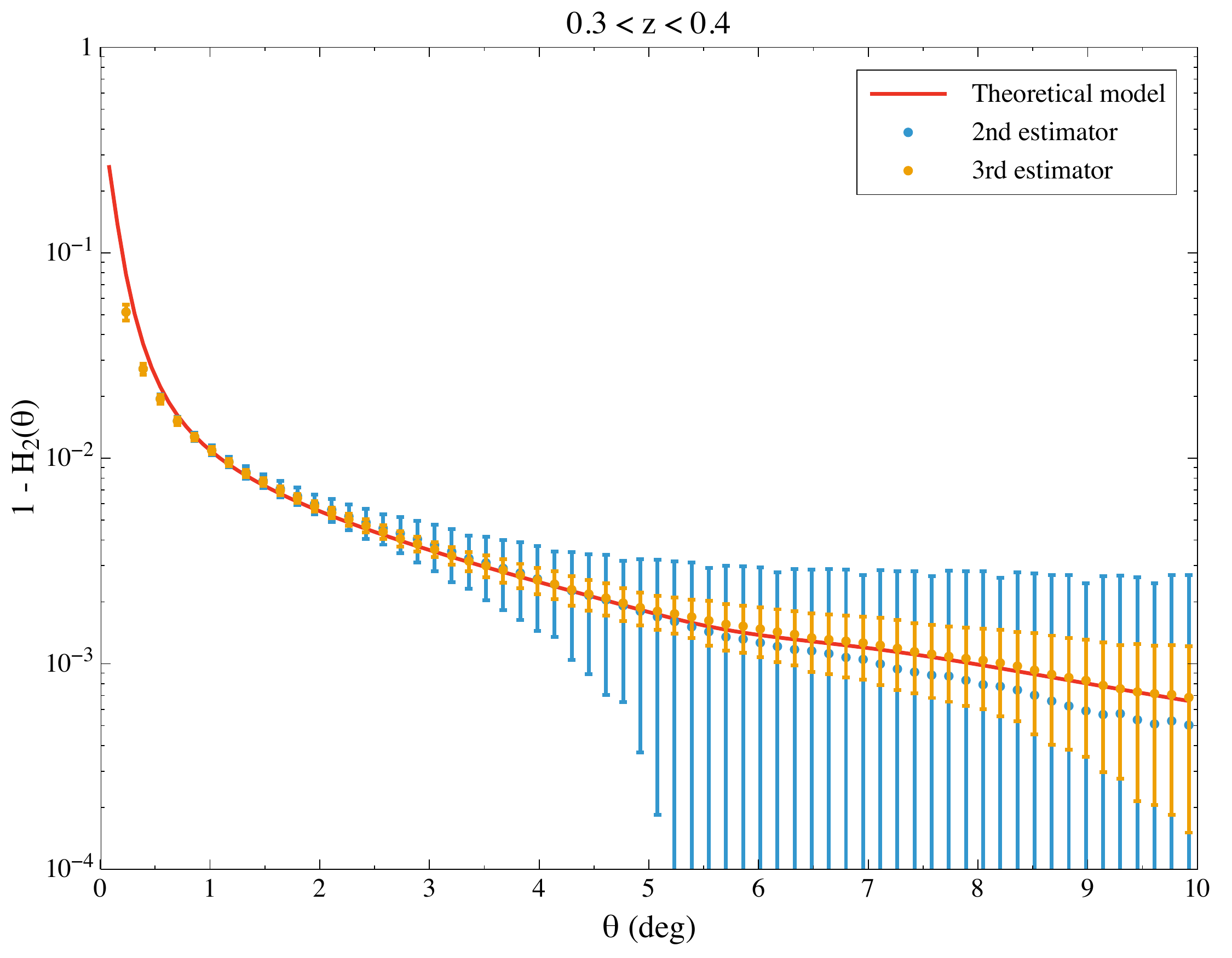}
          \includegraphics[width=0.3\textwidth]{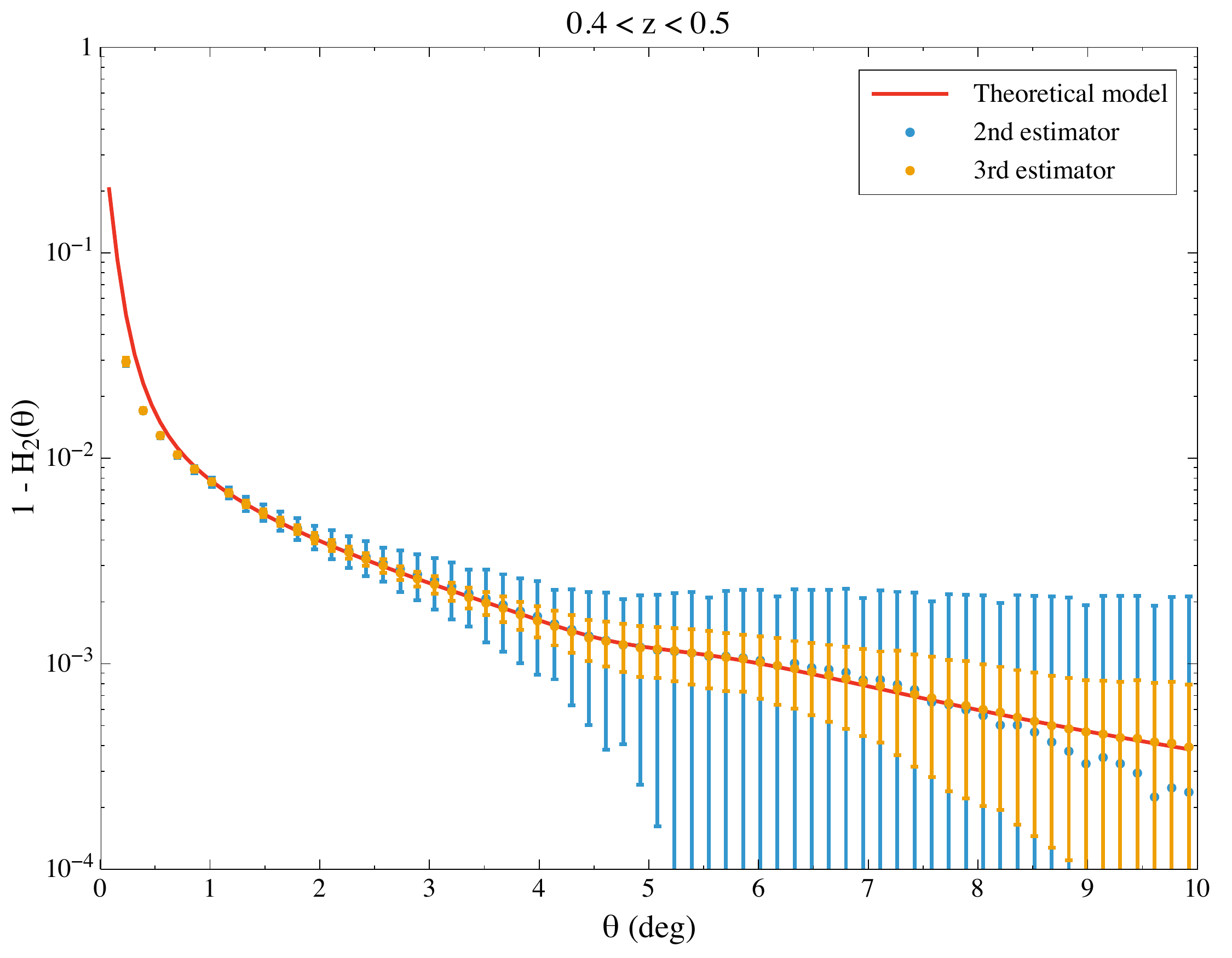}
          \includegraphics[width=0.3\textwidth]{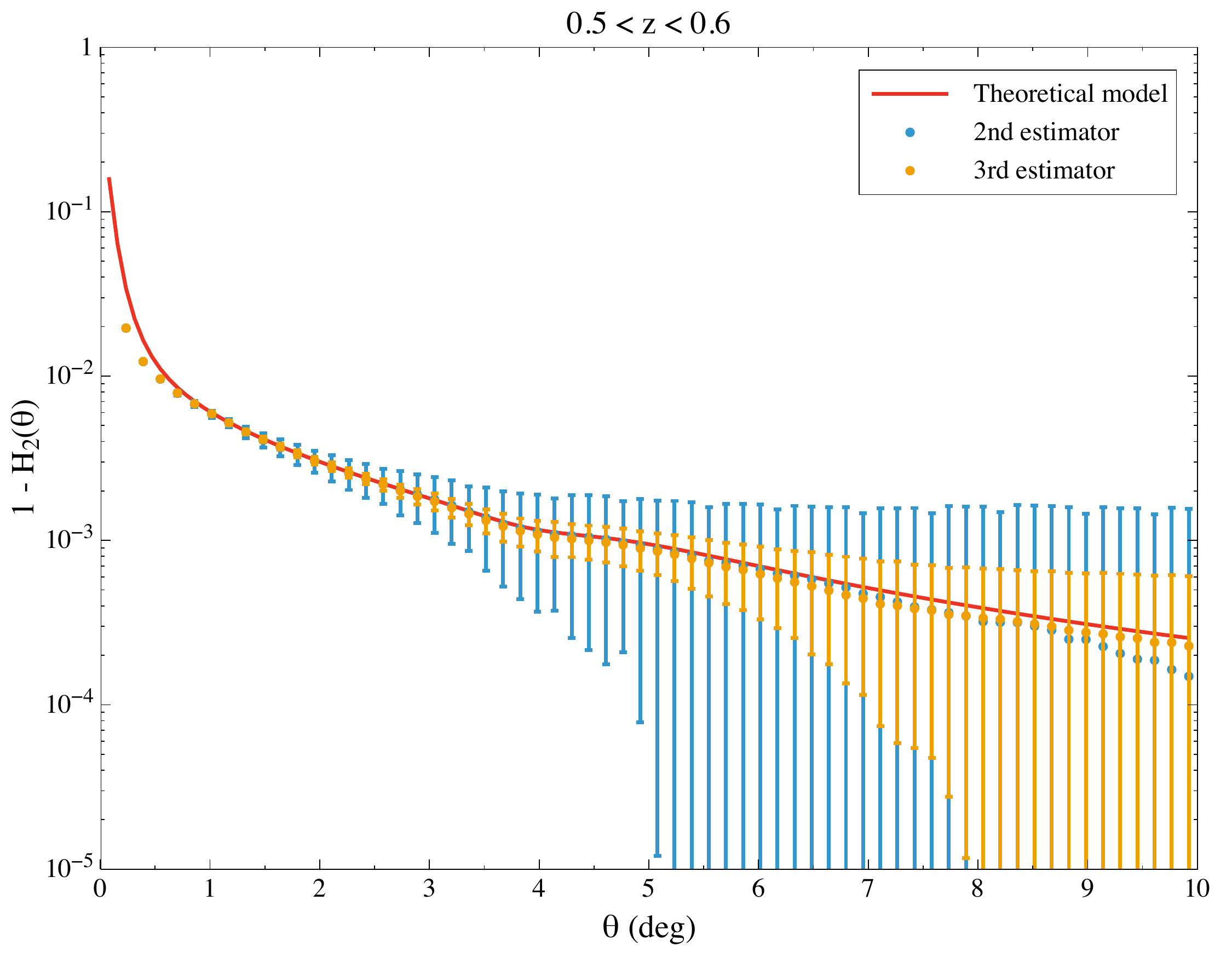}
          \includegraphics[width=0.3\textwidth]{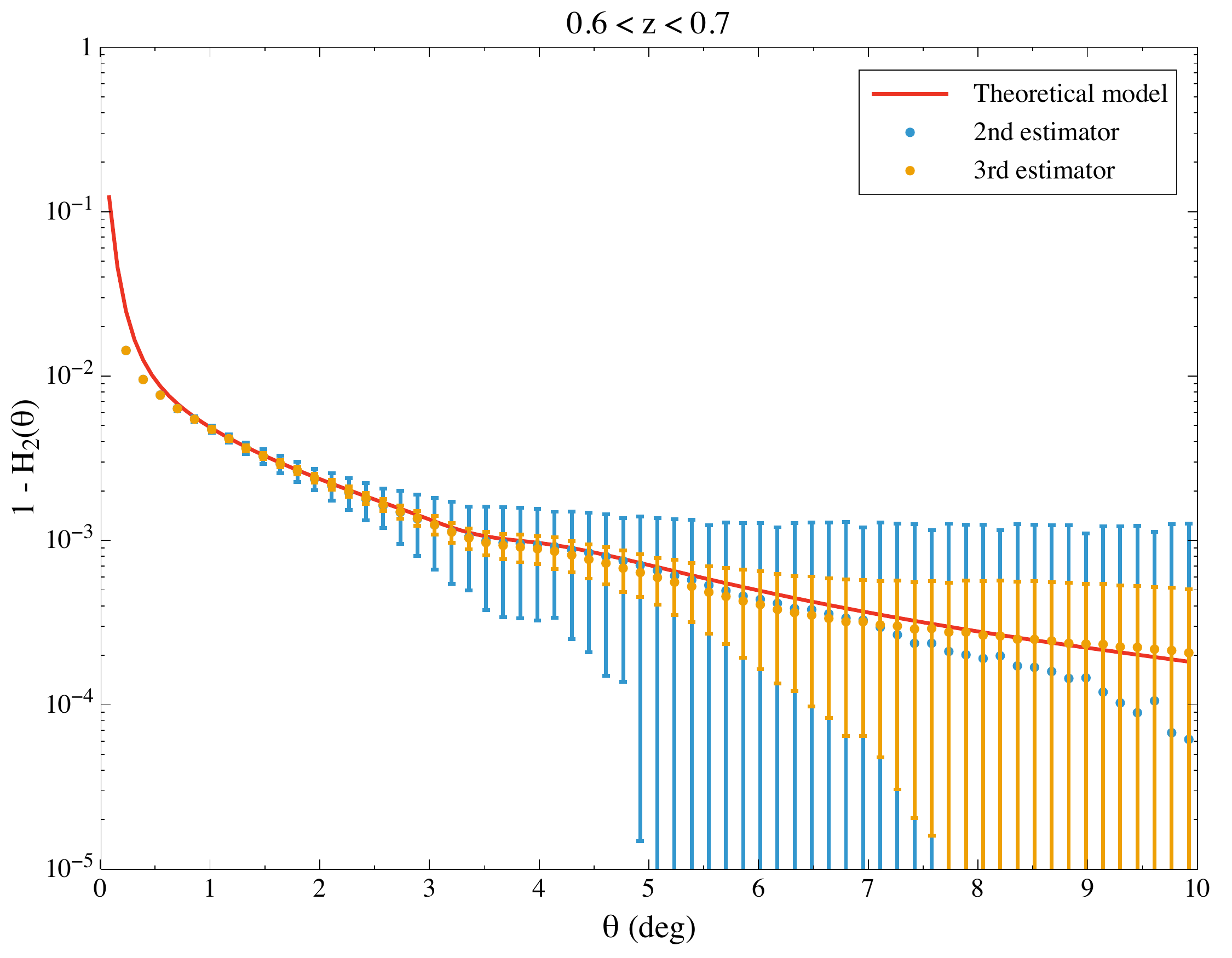}
          \includegraphics[width=0.3\textwidth]{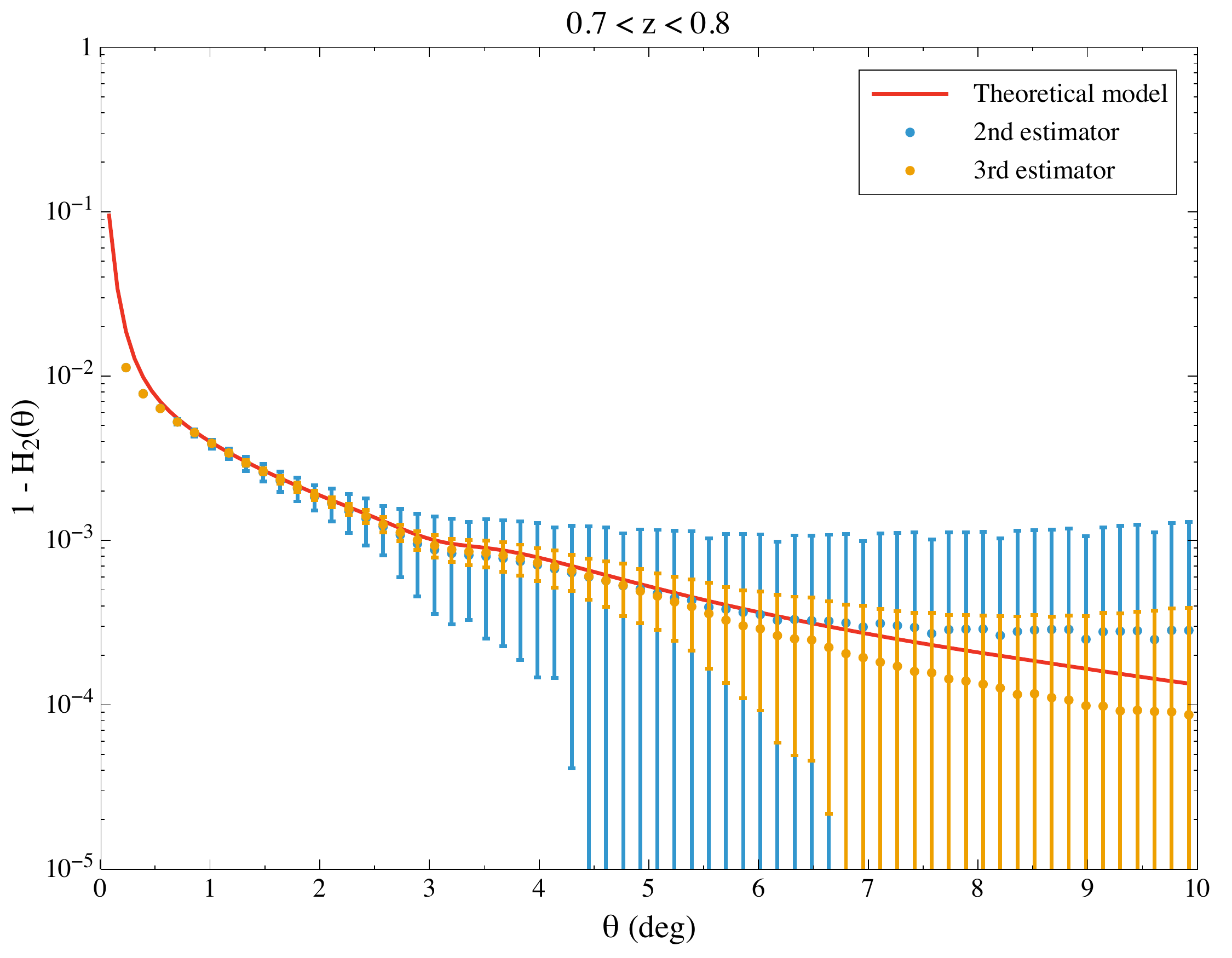}
          \includegraphics[width=0.3\textwidth]{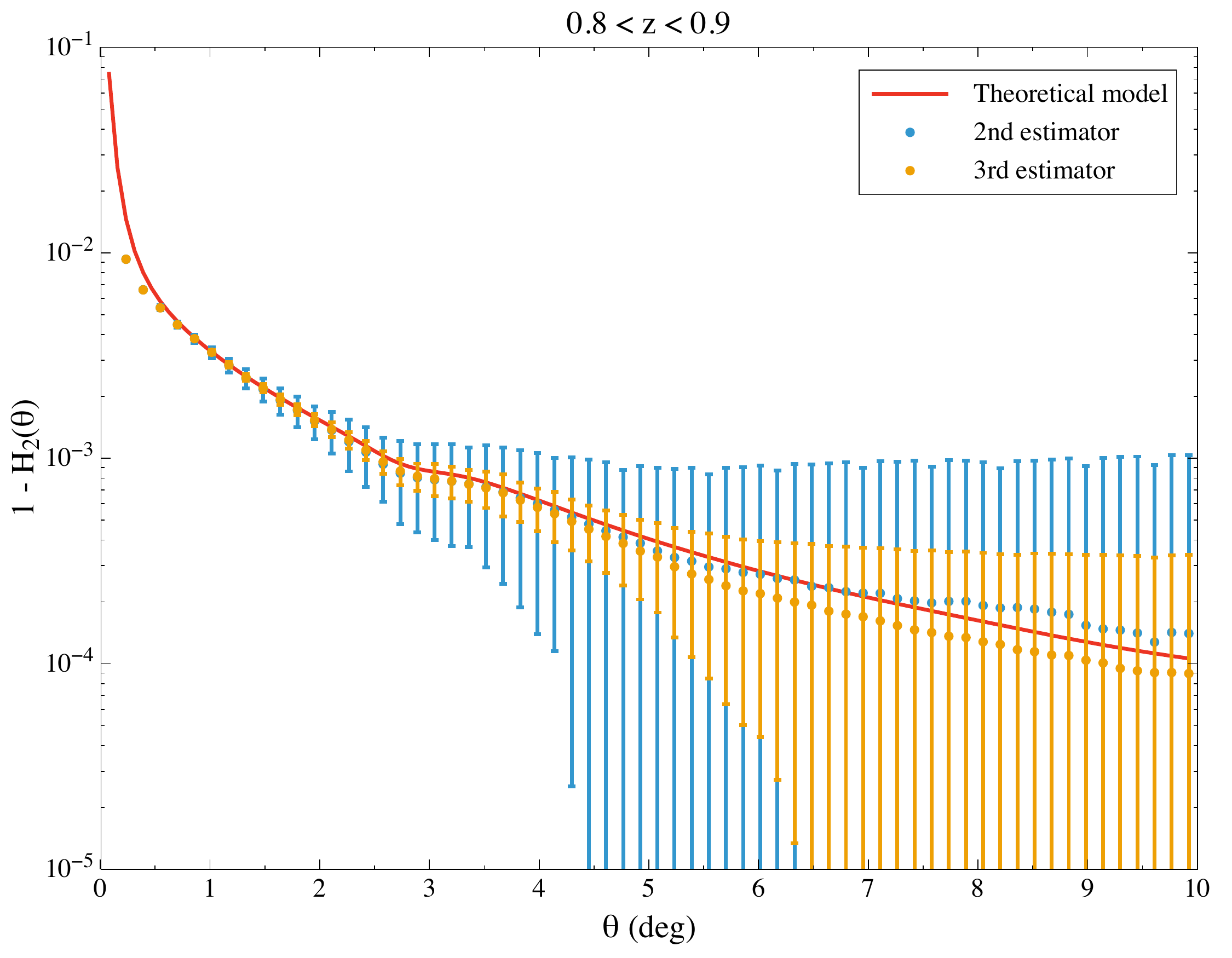}
          \includegraphics[width=0.3\textwidth]{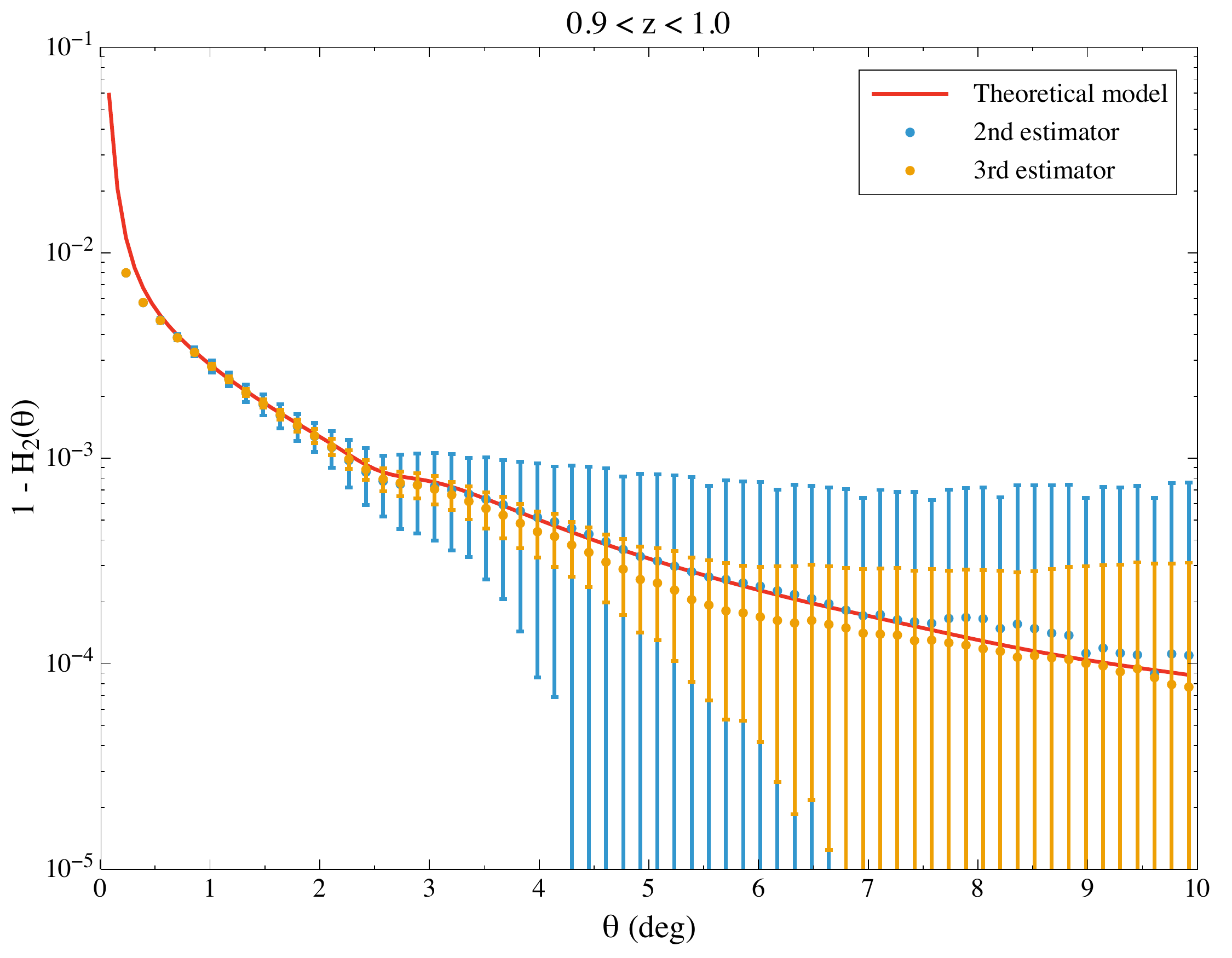}
          \includegraphics[width=0.3\textwidth]{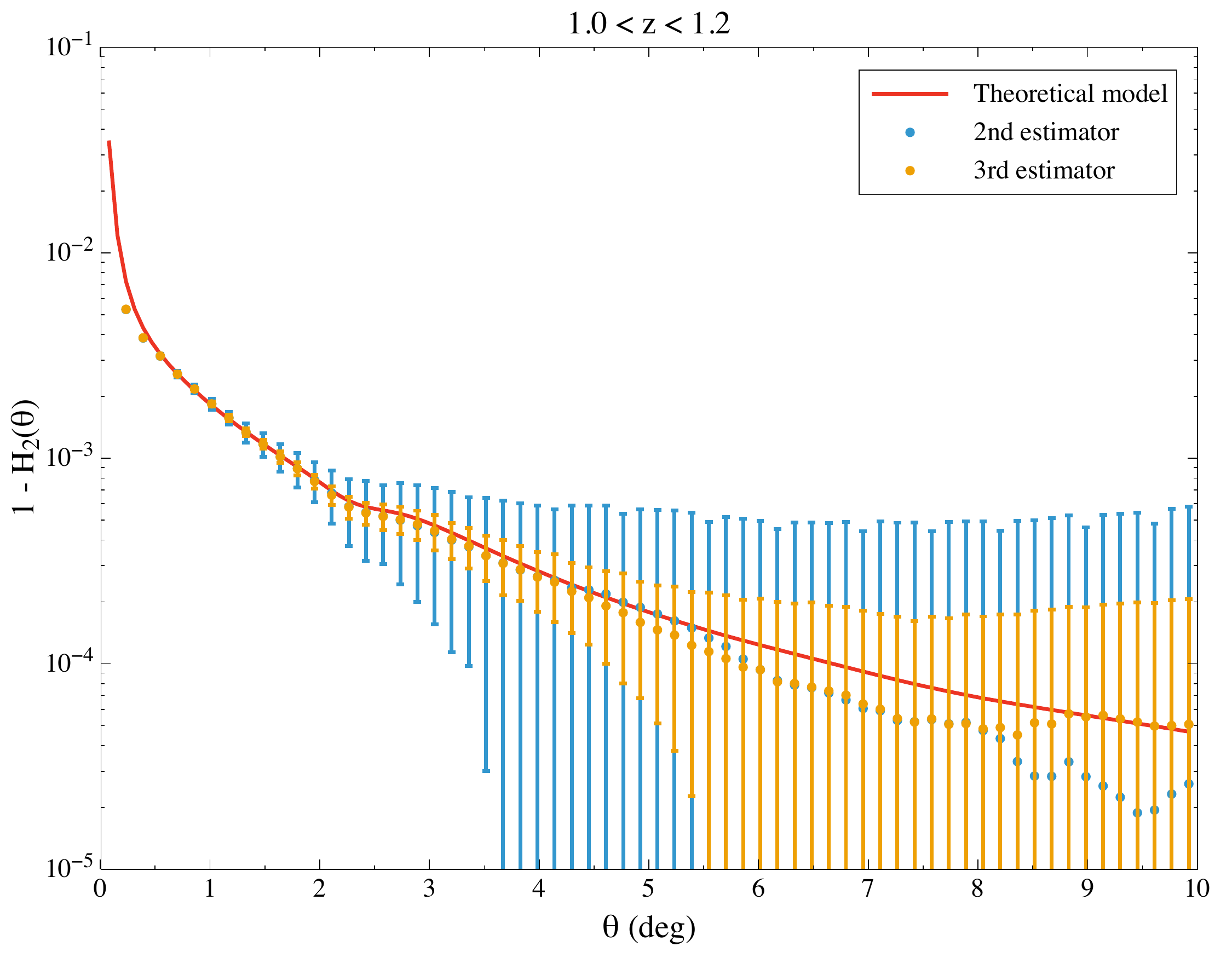}
          \caption{$\Delta H_2(\theta)$ as a function of $\theta$ calculated from the 100 lognormal
                   realizations forthe 9 redshift bins given in table \ref{tab:thetaH}. The data
                   contains Gaussian photometric errors with $\sigma_z=0.03\,(1+z)$. The blue dots
                   with error bars correspond to the mean and standard deviation of the
                   100 mocks for the estimator {\bf E2}, while the red dots correspond to
                   estimator {\bf E3}. The solid red line shows the theoretical model
                   described in section \ref{ssec:modelling_htheta}.}
          \label{fig:bins}
        \end{figure*}
        In order to better understand the approach to homogeneity of a projected galaxy survey
        we have computed $H_2(\theta)$ from the 100 lognormal catalogs using the two estimators
        {\bf E2} and {\bf E3}. Then, the lower limit on the angular homogeneity scale was
        estimated, as described in section \ref{ssec:define_homogeneity}, as the angle for
        which $H_2$ is away from 1 at 95\% C.L. (i.e. $\Delta H_2=1.96\,\sigma_{H_2}$), where
        the errors $\sigma_{H_2}$ were calculated as the standard deviation of the 100 lognormal
        realizations (see section \ref{sssec:errors}).
        
        The comoving three-dimensional homogeneity scale is related to the angular scale $\theta_H$
        through
        \begin{equation}
          r_H(z) \equiv (1+z)\,d_A(z)\,\theta_H(z),
        \end{equation}
        where $d_A(z)$ is the angular diameter distance to redshift $z$.
        
        These results are summarized in table \ref{tab:thetaH}, and can be visualized in figure
        \ref{fig:bins}. The numbers given in this table for $\theta_H$ correspond to the mean
        value obtained from the 100 lognormal mocks, and the errors correspond to the standard
        deviation. Two main observations must be made:
        \begin{table}
          \centering
          \begin{tabular}{|c|cc|cc|}
            \hline
            & \multicolumn{2}{c}{E2} & \multicolumn{2}{|c|}{E3}\\
            \hline
            Bin limits & $\theta_H>$ & $r_H>$ & 
                         $\theta_H>$ & $r_H>$ \\
            \hline
            $0.2 - 0.3$   & $4.9$ & $60$  & $7.6$ & $93$  \\
            $0.3 - 0.4$   & $4.4$ & $74$  & $7.9$ & $131$ \\
            $0.4 - 0.5$   & $4.4$ & $92$  & $7.1$ & $149$ \\
            $0.5 - 0.6$   & $4.0$ & $99$  & $6.4$ & $159$ \\
            $0.6 - 0.7$   & $3.7$ & $105$ & $6.1$ & $175$ \\
            $0.7 - 0.8$   & $3.3$ & $108$ & $5.7$ & $184$ \\
            $0.8 - 0.9$   & $3.4$ & $121$ & $5.4$ & $193$ \\
            $0.9 - 1.0$   & $3.5$ & $137$ & $5.3$ & $207$ \\
            $1.0 - 1.2$   & $3.0$ & $131$ & $4.8$ & $208$ \\
            \hline
          \end{tabular}
          \caption{Lower bound on the scale of homogeneity calculated for the nine redshift 
                   bins of the 100
                   lognormal realizations. The angular scale of homogeneity $\theta_H$
                   is given in degrees, while the corresponding comoving distance
                   is given in ${\rm Mpc}/h$.}\label{tab:thetaH}
        \end{table}
        
        \begin{itemize}
          \item First, since the two estimators make a different use of the data, they have
                different variances, and therefore each of them measures a different lower bound
                on the homogeneity scale. While all the galaxies in the survey are used as
                centres for spherical caps of any angular aperture in the case of {\bf E3},
                only those caps that fit fully inside the field of view are used for {\bf E2}.
                Thus, in this case the variance will grow faster for larger scales, and homogeneity
                is reached on smaller angles. This is explicitly illustrated in figure
                \ref{fig:diag_errors}.
          \item Secondly, the comoving scale corresponding to the angular homogeneity scale for
                each bin seems to increase with redshift. This result is precisely the opposite of
                what intuition would predict: since the amplitude of matter perturbations decreases
                with redshift, the matter distribution is more homogeneous at earlier times, and
                should reach homogeneity on smaller scales at larger redshifts. This paradox is due
                to the fact that the definition that we have used for the scale of homogeneity is
                based on statistical principles, and not on the physical meaning of homogeneity.
                For this and other reasons we believe that producing a number for $\theta_H$ or
                $r_H$ is not as relevant as setting a lower limit to the departure from large-scale
                homogeneity that can be allowed given our observational capabilities.                
        \end{itemize}

      \subsubsection{Statistical uncertainties}
      \label{sssec:errors}
        \begin{figure*}
          \centering
          \includegraphics[width=0.45\textwidth]{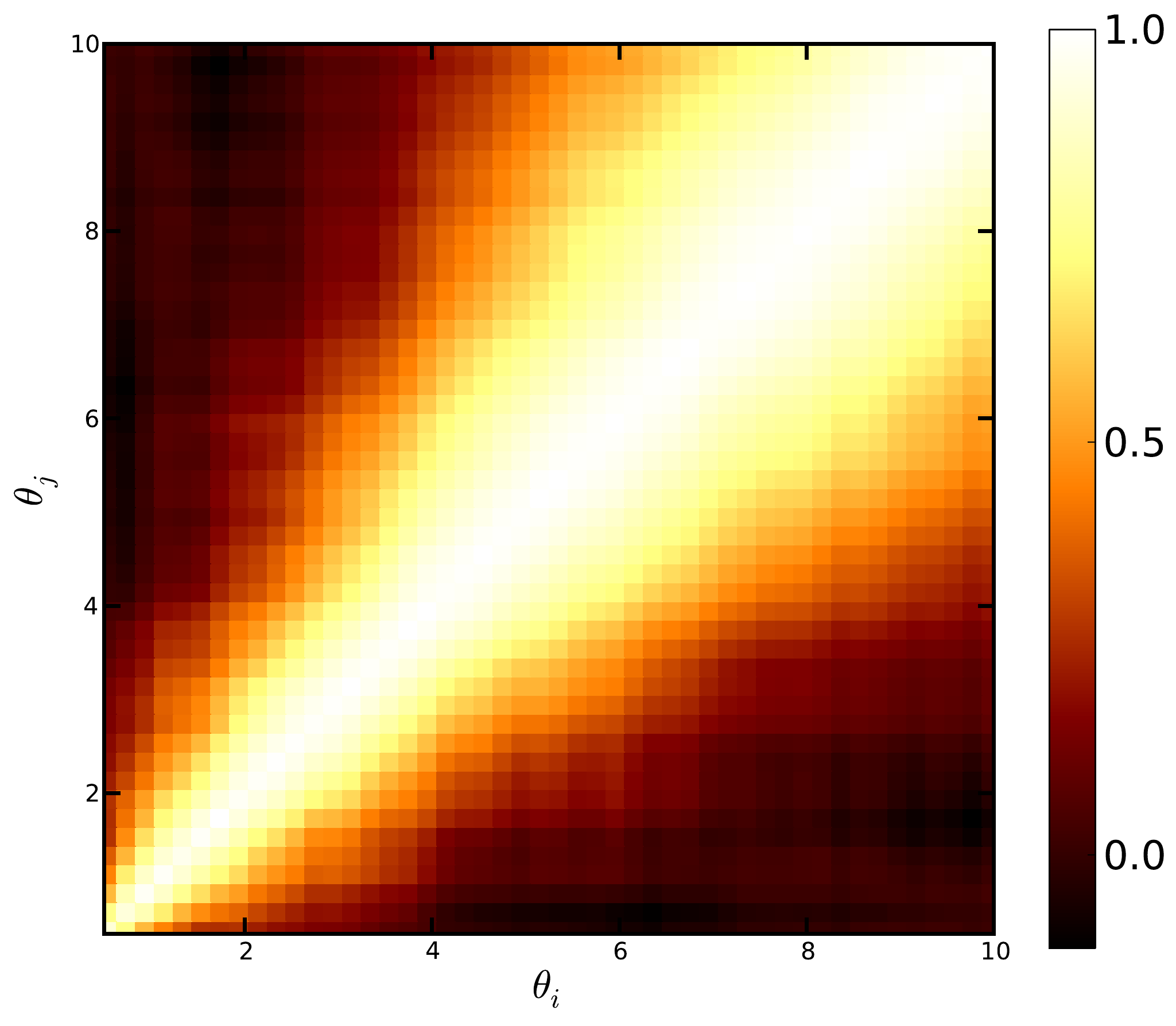}
          \includegraphics[width=0.45\textwidth]{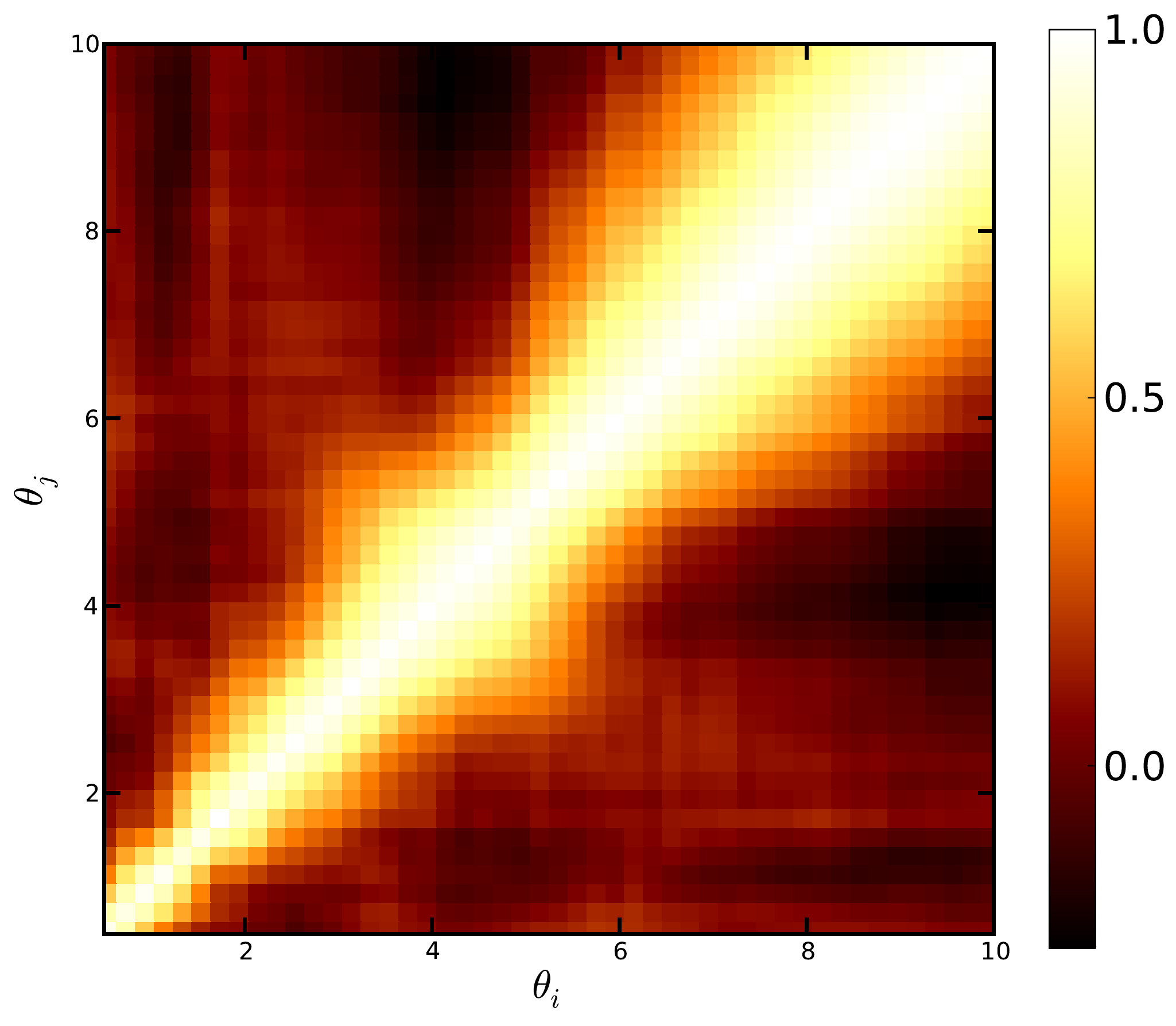}
          \caption{Correlation matrix $\rho_{i,j}\equiv C_{i,j}/\sqrt{C_{i,i}C_{j,j}}$ of
                   $H_2$ for {\bf E2} (left) and {\bf E3} (right). Note that the errors are
                   correlated over relatively wide ranges of scales, specially on large
                   angles.}
          \label{fig:covariances}
        \end{figure*}
        We have studied the full covariance matrix of the angular homogeneity index $H_2(\theta)$
        for the different estimators. The covariance between the angular bins $\theta_i$ and
        $\theta_j$ is calculated from the measurements of $H_2$ in the 100 lognormal mock
        catalogs as
        \begin{equation}
          C_{i,j} = \frac{1}{N_m-1}\sum_{n=1}^{N_m}
                                       H_2^n(\theta_i)H_2^n(\theta_j)-
                                       \overline{H_2(\theta_i)}\,\overline{H_2(\theta_j)},
        \end{equation}
        where $N_m=100$, $H_2^n$ is the measurement on the $n$-th catalog and $\overline{H_2}$ is
        the arithmetic mean over all the catalogs.
        
\begin{figure}
          \centering
          \includegraphics[width=0.45\textwidth]{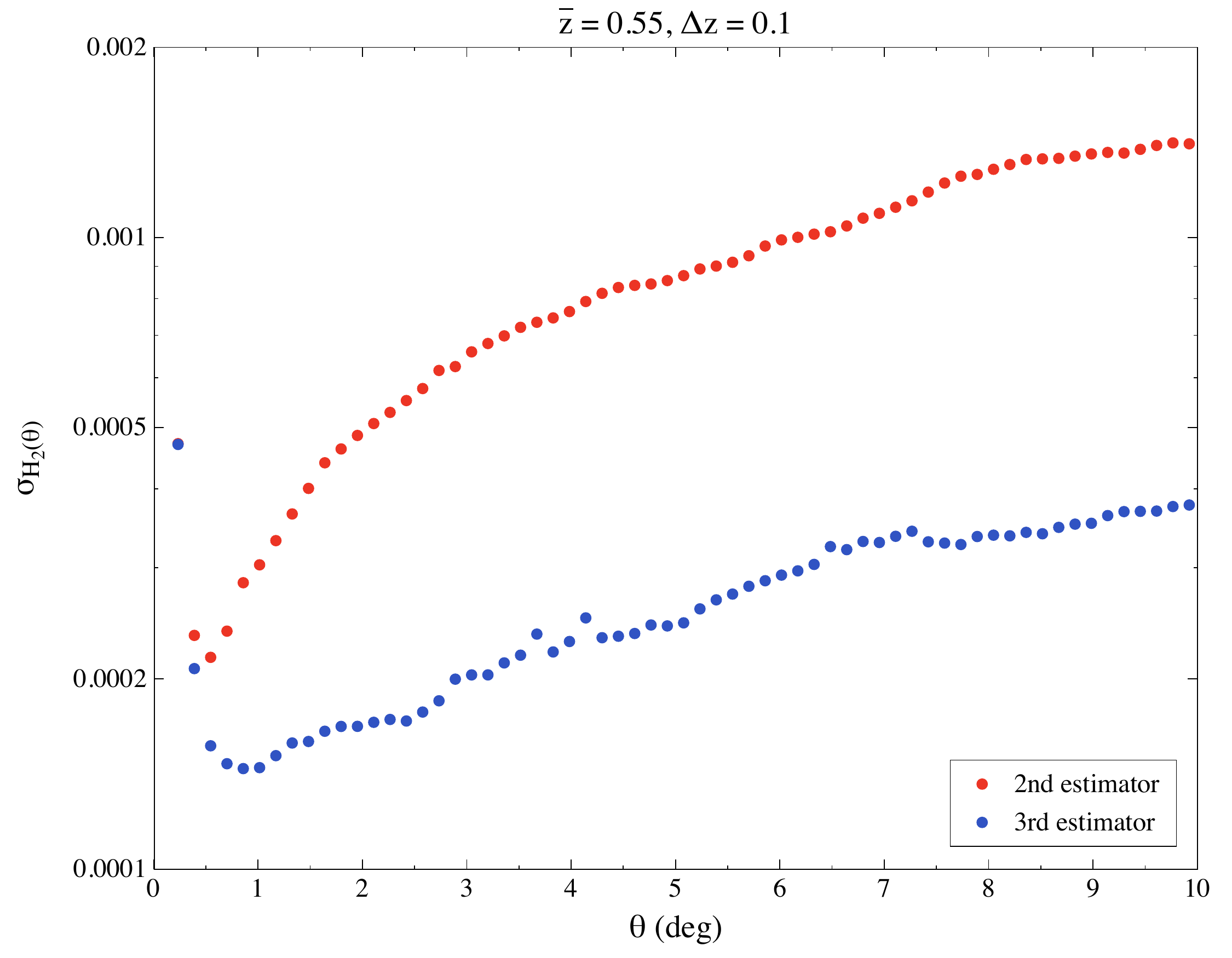}
          \caption{Diagonal errors on $H_2(\theta)$ as a function of $\theta$ for estimators
                   {\bf E2} (red) and {\bf E3} (blue). Since with {\bf E3} all galaxies are
                   used as centres of spherical caps for all $\theta$, the errors are
                   significantly smaller than in the case of {\bf E2}.}
          \label{fig:diag_errors}
        \end{figure}

        Figure \ref{fig:diag_errors} shows the diagonal errors $\sigma_i\equiv\sqrt{C_{i,i}}$ for
        the bin $z\in(0.5,0.6)$ using the estimators {\bf E2} and {\bf E3}. As was noted before,
        the errors corresponding to {\bf E2} are significantly larger than those of {\bf E3} for
        large scales, due to the smaller number of galaxies used as centres of spherical caps for
        those angles.
        
        The correlation matrix $\rho_{i,j}\equiv C_{i,j}/\sqrt{C_{i,i}C_{j,j}}$ is shown, for the
        same two estimators and the same bin, in figure \ref{fig:covariances}. As is shown in the
        figure, the measurements of $H_2$ are statistically correlated over wider ranges of
        scales as we go to larger angles, especially in the case of {\bf E2}. Therefore, if any
        likelihood analysis is to be done on $H_2$, the full covariance matrix must be used.
        
      \subsubsection{Projection effects}
        \begin{figure}
          \centering
          \includegraphics[width=0.45\textwidth]{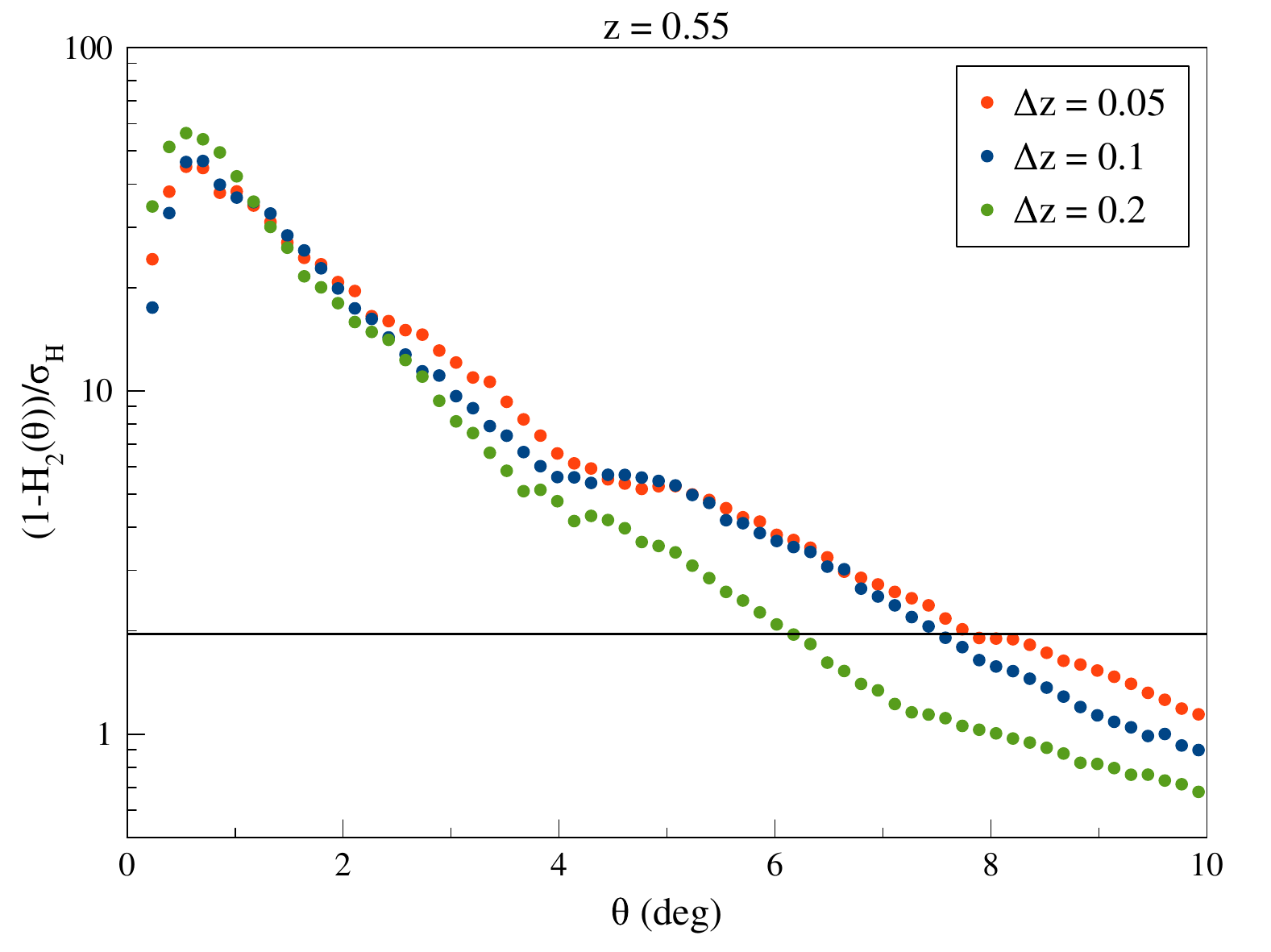}
          \caption{Departure from homogeneity $\Delta H_2$ divided by the uncertainty on
                   $\sigma_{H_2}$ as a function of $\theta$ for a redshift bin centered on
                   $\bar{z}=0.55$ with different binwidths. As could be expected, projecting on
                   wider bins moves the scale of homogeneity towards smaller angles.}
          \label{fig:bins_ln}
        \end{figure}
        As has been said before, using wider redshift bins damps the amplitude of the correlation
        function and makes the projected galaxy distribution more homogeneous (i.e., $H_2$ gets
        closer to 1). However, the amplitude of the error on the correlation function (or on $H_2$)
        will also be damped, and it is therefore interesting to study whether the two dampings 
        compensate each other and to quantify the effect on the scale of homogeneity. This has
        been done in figure \ref{fig:bins_ln}. The homogeneity index $H_2(\theta)$ has been
        calculated at $\bar{z}=0.55$ using different redshift binwidths: $\Delta z = 0.05,\,
        0.1,\,0.2$, and the ratio $\Delta H_2/\sigma_{H_2}$ has been plotted for different values
        of $\theta$. According to our definition, the homogeneity scale is reached when this ratio 
        becomes 1.96. As can be seen, the damping of $\Delta H_2$ due to projection effects is not
        compensated by the corresponding damping on $\sigma_{H_2}$, and the homogeneous regime is
        reached on smaller scales for wider bins, as could be intuitively expected. Since the use
        of photometric redshifts effectively increases the width of the redshift bin, it produces
        a similar effect.
        
      \subsubsection{Fraction of the sky}\label{sssec:fsky}
        \begin{figure}
          \centering
          \includegraphics[width=0.45\textwidth]{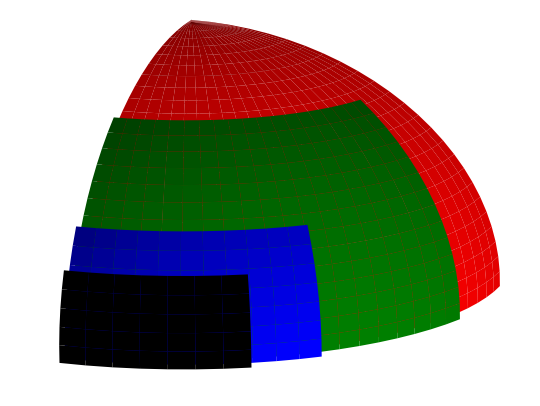}
          \caption{The four survey cases considered in section \ref{sssec:fsky}, covering
                   $5000\,{\rm deg}^2$ (red), $3000\,{\rm deg}^2$ (green), $1000\,{\rm deg}^2$
                   (blue) and $500\,{\rm deg}^2$ (black).}\label{fig:regions}
        \end{figure}
        \begin{figure*}
          \centering
          \includegraphics[width=0.45\textwidth]{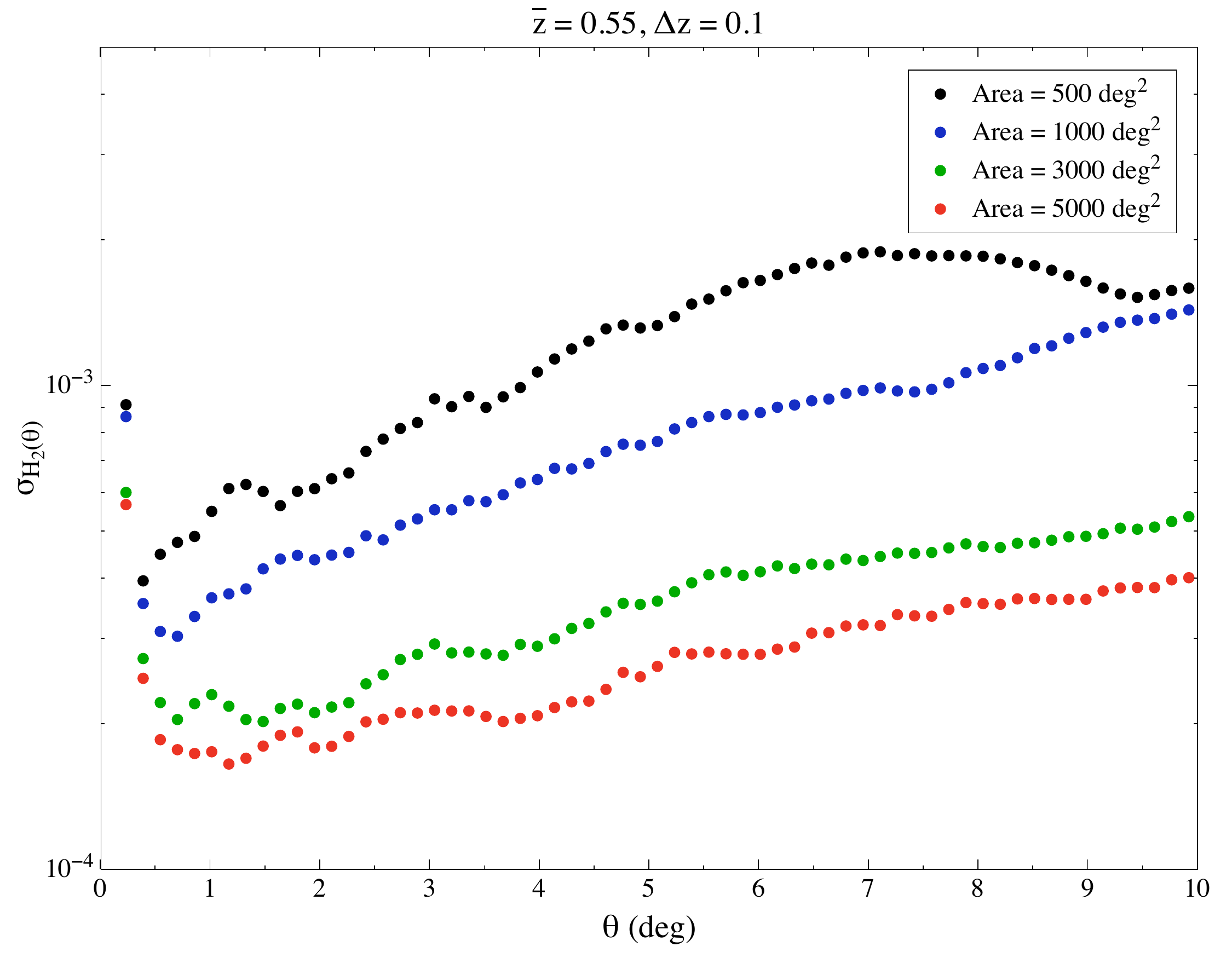}
          \includegraphics[width=0.45\textwidth]{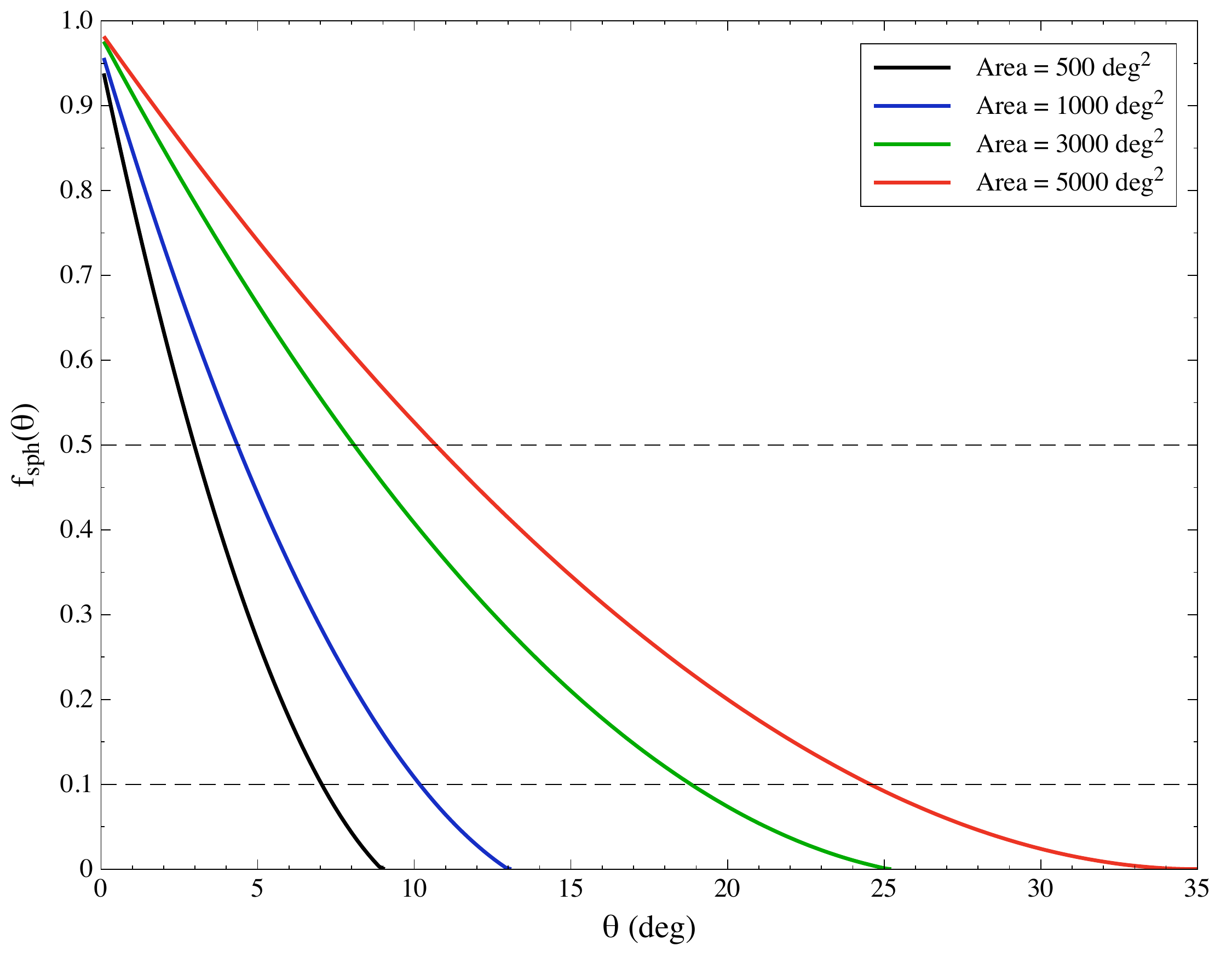}
          \caption{{\sl Left panel}: Errors on $H_2(\theta)$ for the four surveys described
                   in figure \ref{fig:regions} as a function of $\theta$. {\sl Right panel}:
                   Fraction of complete spheres of different radii in the same four surveys
                   as a function of $\theta$. The horizontal dashed lines correspond to the
                   last two criteria considered below to determine the maximum angular scale
                   to be used.}
          \label{fig:error_areas}
        \end{figure*}

        \begin{table*}
          \centering
          \begin{tabular}{|c|cc|cc|cc|}
            \hline
            & \multicolumn{2}{c|}{Crit. 1}   &
              \multicolumn{2}{c|}{Crit. 2} &
              \multicolumn{2}{c|}{Crit. 3} \\
            \hline
            Area & $\theta_{\rm max}$ & $\Delta H_2^{\rm min}\times10^3$ & 
                   $\theta_{\rm max}$ & $\Delta H_2^{\rm min}\times10^3$ &
                   $\theta_{\rm max}$ & $\Delta H_2^{\rm min}\times10^3$ \\
            \hline
            $5000\,{\rm deg}^2$   & $35^o $ & $0.08 \pm 1   $
                                  & $25^o $ & $0.08 \pm 0.8 $
                                  & $10^o $ & $0.2  \pm 0.35$\\
            $3000\,{\rm deg}^2$   & $25^o $ & $0.08 \pm 1.5 $
                                  & $19^o $ & $0.09 \pm 0.9 $
                                  & $8^o  $ & $0.4  \pm 0.5 $\\
            $1000\,{\rm deg}^2$   & $13^o $ & $0.12 \pm 1.7 $
                                  & $10^o $ & $0.2  \pm 1.4 $
                                  & $4.5^o$ & $1.0  \pm 0.7 $\\
            $ 500\,{\rm deg}^2$   & $9^o  $ & $0.3  \pm 2   $
                                  & $7^o  $ & $0.33 \pm 1.7 $
                                  & $3^o  $ & $1.8  \pm 0.8 $\\
            \hline
          \end{tabular}
          \caption{Constraints on the level of homogeneity $\Delta H_2^{\rm min}\equiv
                   1-H_2(\theta_{\rm max})$ attainable with surveys of different areas
                   for the three criteria described in section \ref{sssec:fsky} for the 
                   maximum angular scale $\theta_{\rm max}$.}\label{tab:thetamax}
        \end{table*}
        In order to study the effects related to the area covered by a given survey, we considered a
        fiducial redshift bin $0.5 < z < 0.6$ and restricted the data from
        our mock catalogs to regions of different areas. Specifically, we have considered surveys
        covering $\sim5000\,{\rm deg}^2$ (one octant of the sky) $\sim3000\,{\rm deg}^2$,
        $\sim1000\,{\rm deg}^2$ and $\sim500\,{\rm deg}^2$. For simplicity we have used simply
        connected fields of view with the shapes shown in figure \ref{fig:regions}. This is an
        ideal scenario, and therefore the results shown here would correspond to the most
        optimistic ones any survey of the same area could obtain. The total area covered by a
        given survey affects the measurement of $H_2(\theta)$ in two ways.
        
        First, the sample variance should be inversely proportional to $\sqrt{f_{\rm sky}}$
        \citep{Crocce:2010qi}, and therefore the uncertainty in $H_2$ will grow for smaller areas.
        This is illustrated in the left panel of figure \ref{fig:error_areas}, which shows the
        magnitude of the errors on $H_2$ for the 4 different areas.
        
        Secondly, the survey size limits the maximum scale that we are able to probe, and may
        prevent us from reaching the homogeneous regime. In order to illustrate this point, we
        have performed the following exercise: inside each of the regions shown in figure
        \ref{fig:regions}, we have randomly placed a large number points. Then, for different
        values of $\theta$, we have estimated the fraction of spherical caps of radius $\theta$
        centered on these points that lie fully inside the surveyed region. The result is shown
        in the right panel of figure \ref{fig:error_areas}. In view of this result we have
        established three different criteria to define the largest scale $\theta_{\rm max}$
        that can be probed in a survey:
        \begin{enumerate}
          \item $\theta_{\rm max}$ corresponds to the radius of the largest spherical cap that
                fits inside the surveyed region.
          \item $\theta_{\rm max}$ is the angle for which the fraction shown in the right panel
                of figure \ref{fig:error_areas} is 10\%.
          \item The same as above for a fraction of 50\%.
        \end{enumerate}
        For each of these criteria and for the 4 different areas we have listed in table
        \ref{tab:thetamax} the minimum value of $\Delta H_2$ that can be obtained together with
        its uncertainty $\sigma_{H_2}$.
        
  \section{Robustness of the method}
  \label{sec:robustness}
    Since the use of random catalogs to correct for mask and edge effects may bias the
    estimation of $H_2(\theta)$ towards homogeneity, it is important to study the relevance
    of this effect. Furthermore, it has been argued \citep{1997EL.....40..491D} that fractal
    distributions may look homogeneous when projected onto the celestial sphere, and therefore
    it is necessary to verify that we are indeed able to distinguish a 3D fractal from an
    asymptotically homogeneous distribution using only angular information, and to what level
    so. In order to address these questions, we have analyzed different inhomogeneous models which,
    we know, should not approach homogeneity.
    \begin{figure*}
      \centering
      \includegraphics[width=0.45\textwidth]{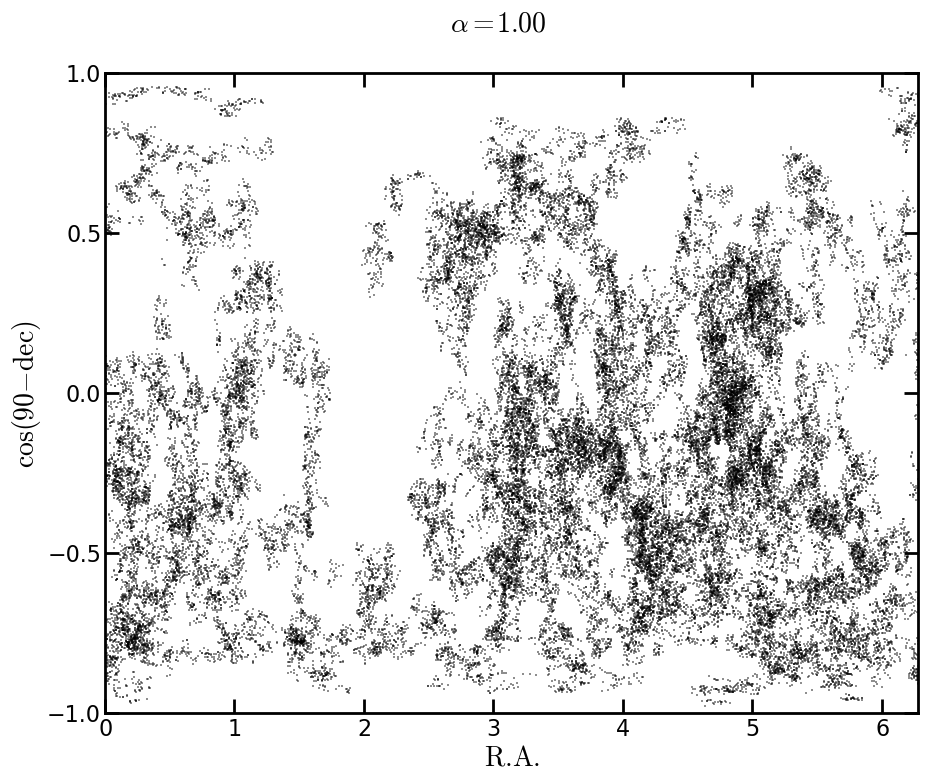}
      \includegraphics[width=0.45\textwidth]{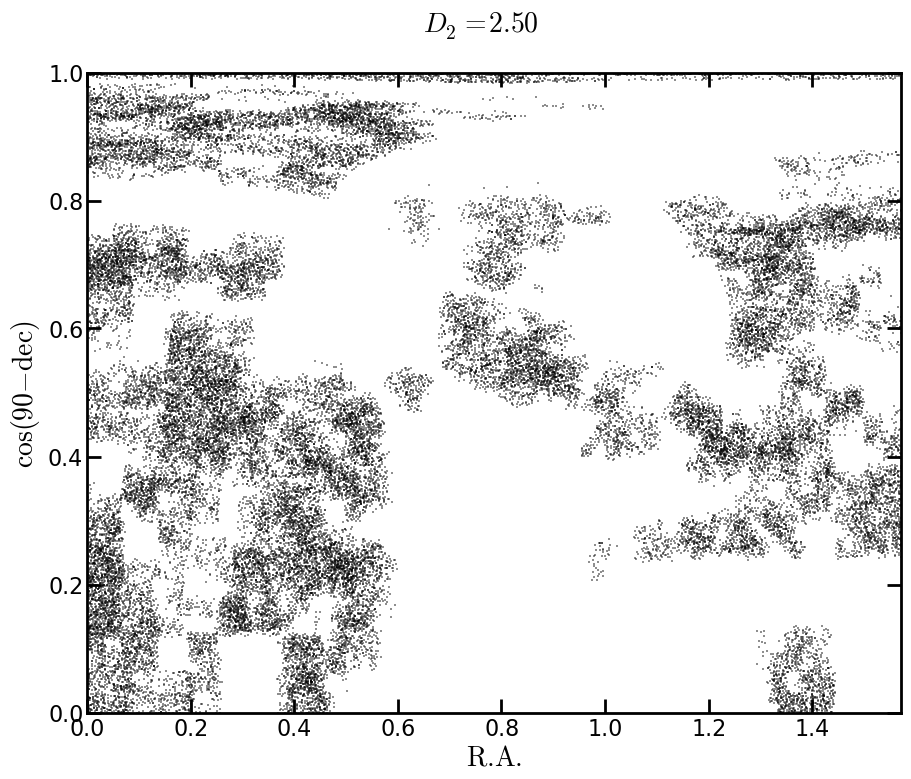}
      \includegraphics[width=0.45\textwidth]{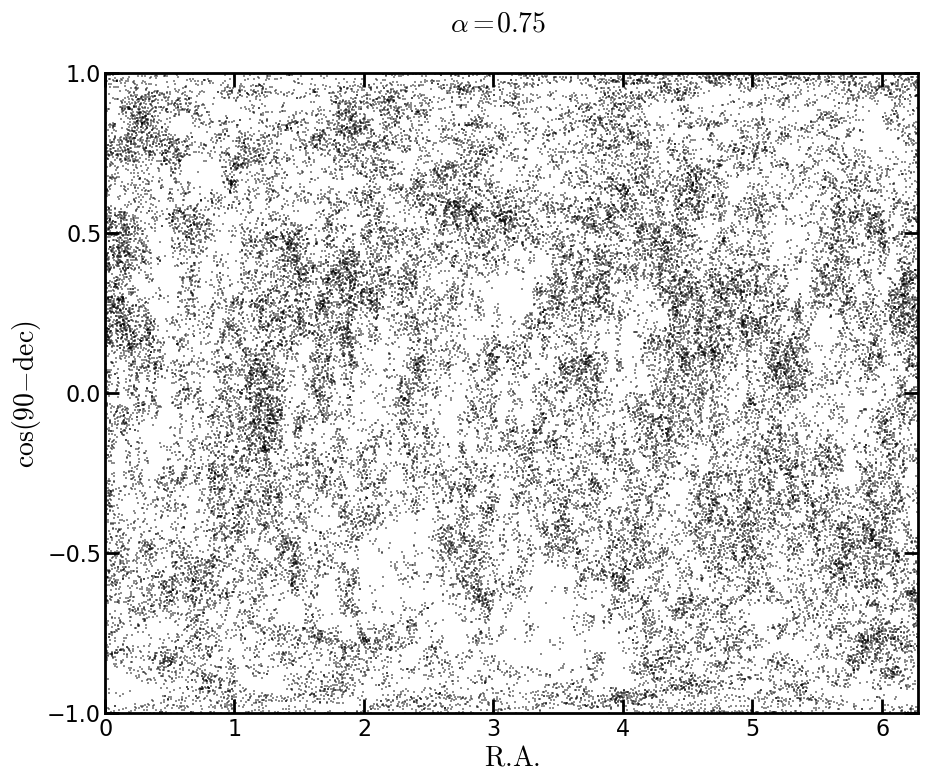}
      \includegraphics[width=0.45\textwidth]{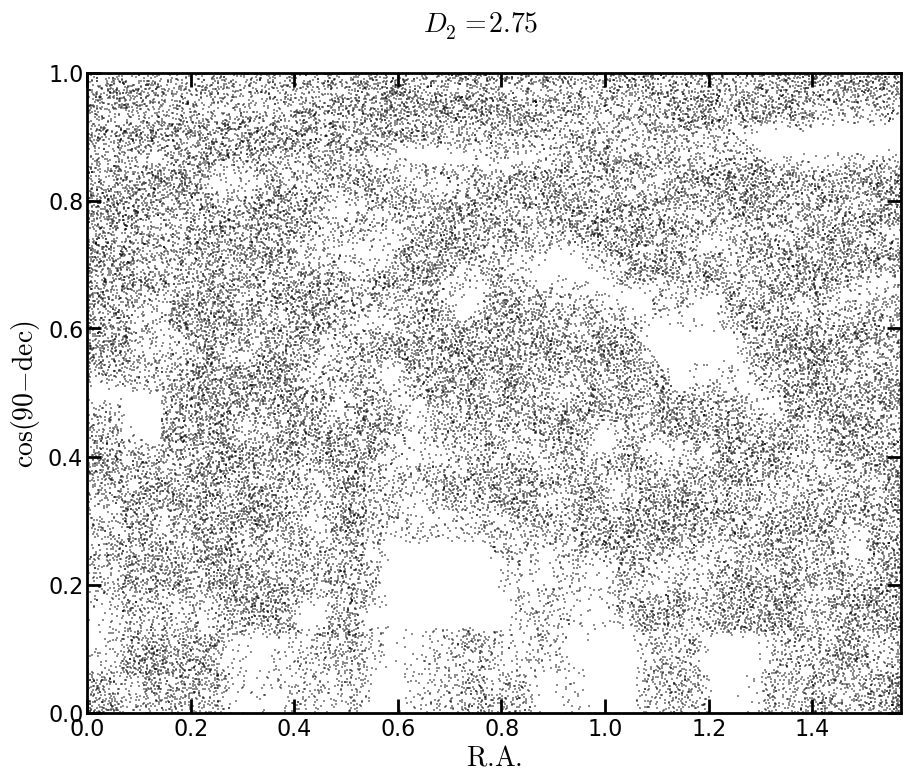}
      \includegraphics[width=0.45\textwidth]{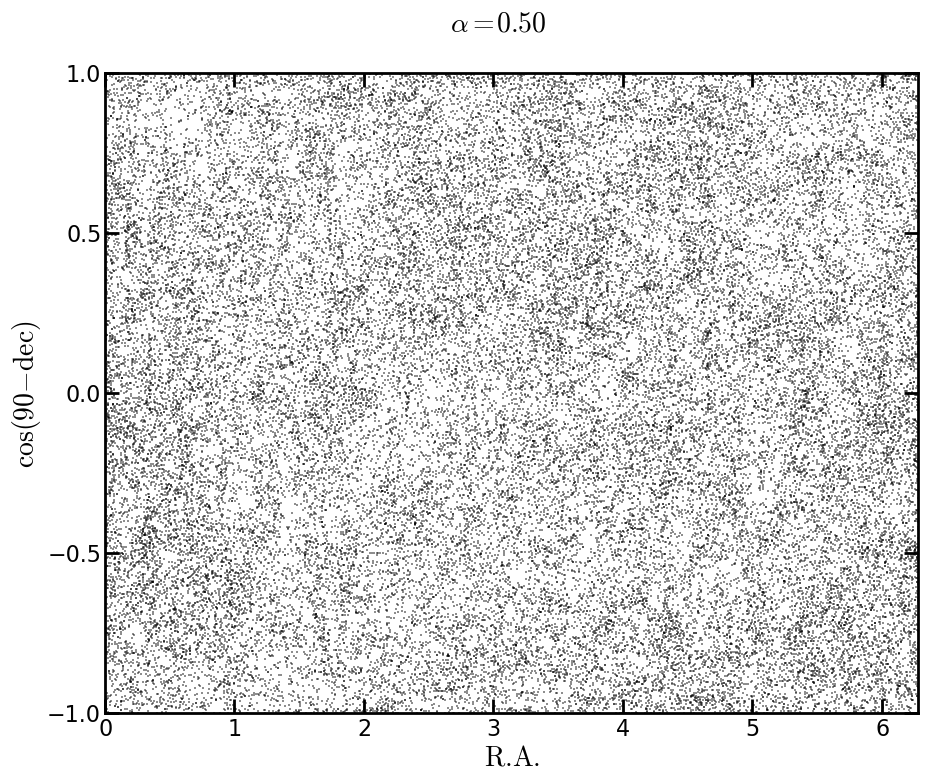}
      \includegraphics[width=0.45\textwidth]{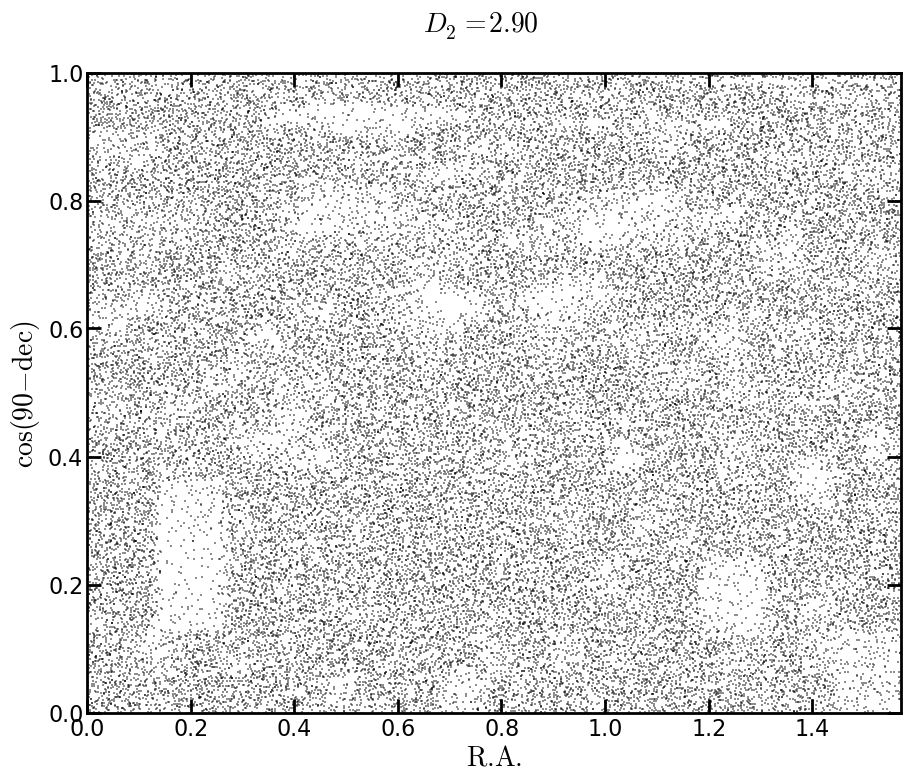}
      \caption{Sky maps for the 2D Rayleigh-Levy flights (left) and the
               $\beta$-model catalogs at $0.5 < z < 0.6$ (right). The
               plots are ordered top-down from more to less inhomogeneous.}
      \label{fig:maps}
    \end{figure*}
    
    \subsection{Spherical Rayleigh-Levy flights}
      A random walk in 3 dimensions is an iterative point process in which the distance between
      one point and the next one is drawn from a probability distribution independently of all
      previous jumps. In the particular case of a heavy-tailed Pareto distribution
      \begin{equation}
        P(r>R)=\left\{\begin{array}{lr}
                        1 & R<R_0\\
                        \left(\frac{R}{R_0}\right)^{-\alpha} & R\geq R_0
                      \end{array}\right.,
      \end{equation}
      these walks are called \emph{L\'evy flights} and exhibit a fractal behavior with
      $D=\alpha$ for $\alpha\leq2$ \citep{2000MNRAS.313L..39N}.
           
      We have generated random walks on the sphere by following a similar process. We
      first choose a starting point on the sphere at random, and draw an angular distance
      $\theta_d$ from a probability distribution. The next point is selected at this
      distance in an arbitrary direction from the first one, and the process is repeated.
      For our walks we have chosen a distribution similar to the one given above in the
      three-dimensional case
      \begin{equation}
        P(\theta_d<\theta)=\left\{\begin{array}{lr}
                                    1 & \theta<\theta_0\\
                                    \left(\frac{1-\cos\theta}
                                    {1-\cos\theta_0}\right)^{-\alpha} & \theta\geq\theta_0
                                  \end{array}\right..
      \end{equation}
      
      It must be noted that with this procedure we are generating an inhomogeneous distribution
      directly in the 2-dimensional sphere, and not projecting a 3-dimensional set. However
      we know for sure that this distribution must asymptotically reach some $H_2(\theta)\neq1$,
      and therefore we can use it to verify that the use of random catalogs does not bias our
      results. To do so we have considered values of $\alpha = 0.5,\, 0.75\,\text{and}\,1$,
      generating 20 random walks containing $10^6$ objects in all cases. Figure \ref{fig:maps}
      (left column) shows the 2D distribution of some of these walks for different values of
      $\alpha$, showing that the degree of ihomogeneity increases with $\alpha$.
      
\begin{figure}
        \centering
        \includegraphics[width=0.45\textwidth]{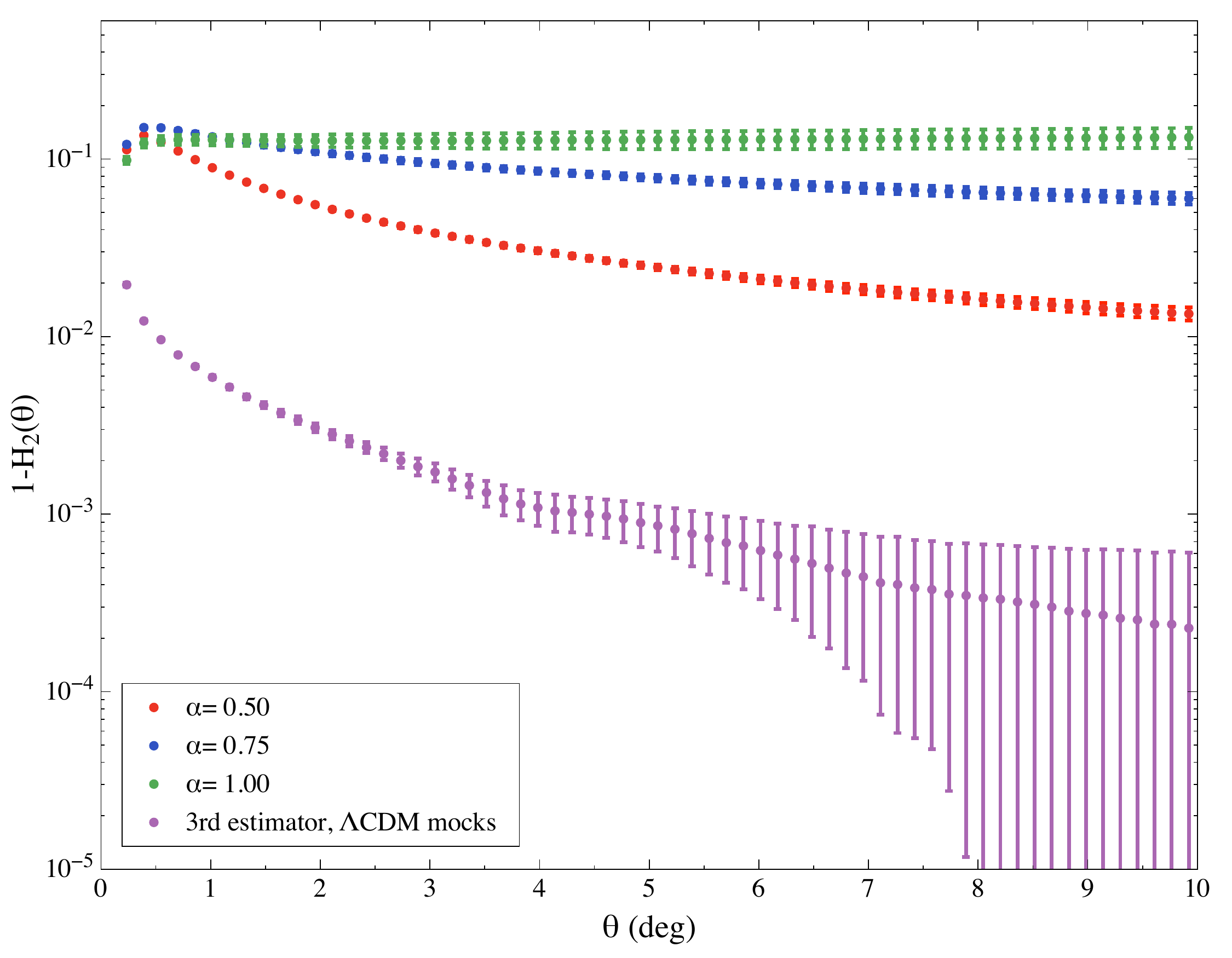}
        \caption{$\Delta H_2(\theta)$ as a function of $\theta$ for three sets of 2D Rayleigh-Levy
                 flights with $\alpha=1.0$ (green), $\alpha=0.75$ (blue) and $\alpha=0.5$ (red),
                 together with the result from the lognormal catalogs for the bin $0.5 < z < 0.6$
                 (purple).}
        \label{fig:hth_2Dflights}
      \end{figure}

      We have calculated $H_2(\theta)$ and its error from these random walks using the {\bf E3}
      estimator. The results are shown in figure \ref{fig:hth_2Dflights} together with those
      corresponding to the $\Lambda$CDM lognormal catalogs. In all cases the asymptotic value of
      $H_2$ is different from 1 and can be clearly distinguished from the $\Lambda$CDM prediction,
      showing that, at least within the range of scales explored, our method is not biased towards
      homogeneity.
      
    \subsection{$\beta$-model}
      \begin{figure}
        \centering
        \includegraphics[width=0.45\textwidth]{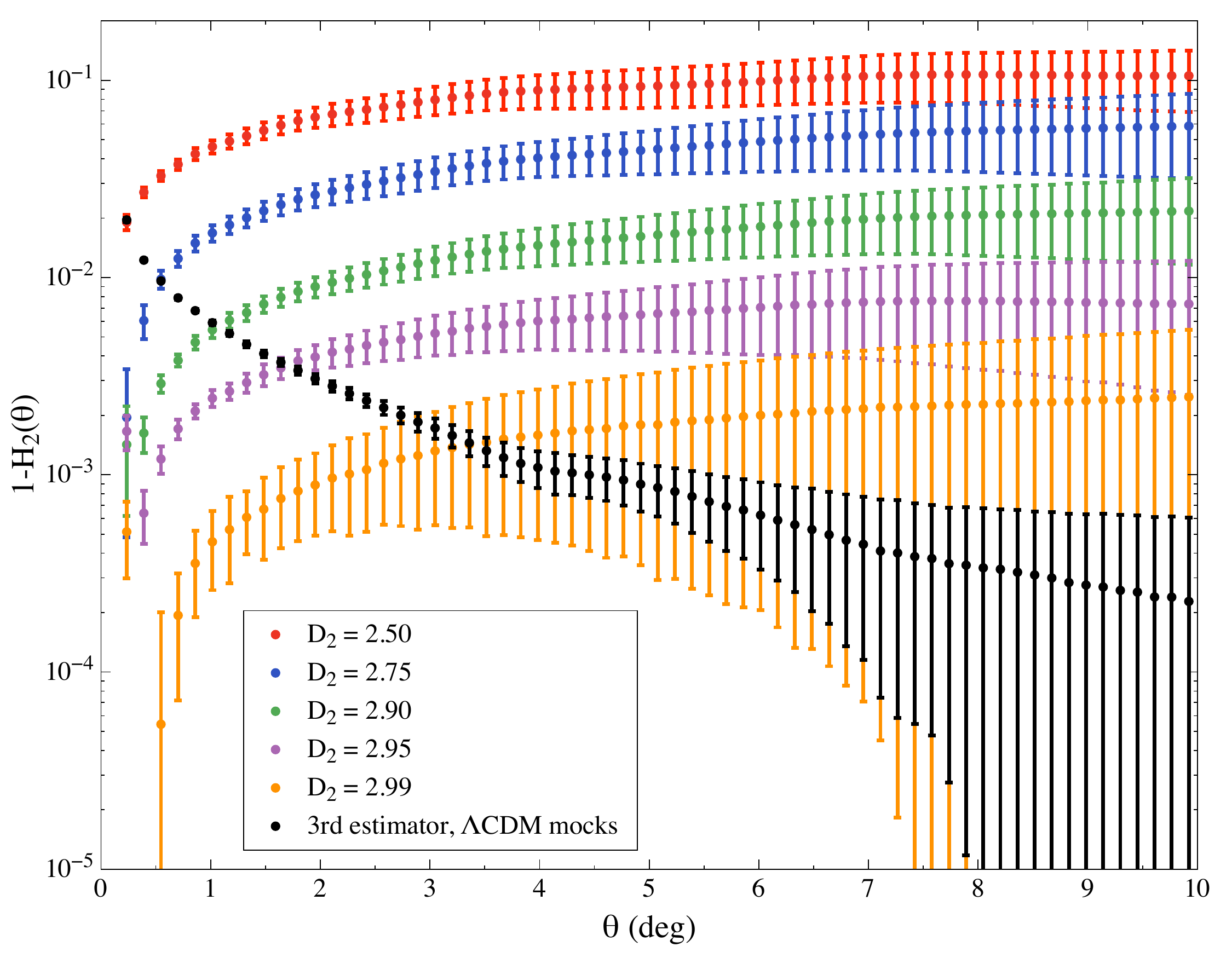}
        \caption{$\Delta H_2(\theta)$ as a function of $\theta$ for the $\beta$-models projected
                 onto the redshift bin $0.5 < z < 0.6$ for different values of $D_2$ from 2.5 (top)
                 to 2.99 (bottom), together with the $\Lambda$CDM prediction obtained
                 from the lognormal mocks (black). In spite of the 2D projection, we are still
                 able to distinguish the inhomogeneous nature of the $\beta$-model from
                 an asymptotically homogeneous model for $D\lesssim 2.95$.}
        \label{fig:hth_beta}
      \end{figure}
      The fractal $\beta$-model (see \citet{1991A&A...246..634C}) is a
      multiplicative cascading fractal model based on the following process: take a cubic box
      of side $L$ and perform $N_{\rm side}$ equal divisions per side. Then, give a
      probability $p<1$ to each of the $N_{\rm side}^3$ sub-cubes of surviving to the next
      iteration and randomly choose those which survive according to this probability.
      In the next iteration you follow the same process on each of the surviving sub-cubes.
      In the $n$-th iteration, the average number of surviving cells will be $N_{\rm surv}=
      (N_{\rm side}^3\,p)^n$. Equating this to $N_{\rm side}^{n\,D}$ we obtain that this set
      has a fractal dimension
      \begin{equation}
        D=3+\log_{N_{\rm side}}p.
      \end{equation}
            
      We explored different values for $D$ ranging from 2.5 to 3, generating multiple realizations
      of this process for each value. These catalogs were produced by running the process outlined
      above on a cubic box of the same size as the one used for the lognormal catalogs, using
      $N_{\rm side}=2$. The catalog is then subsampled to the desired number density and the
      three-dimensional distances to each object are translated into redshifts using our fiducial
      cosmological parameters. This is, of course, not correct, since the distance-redshift
      relation for this model need not be that of FRW, however it is not clear which
      relation should be used. In any case, our aim is to explore whether a three-dimensional
      inhomogeneous model could be noticed when projected onto the sphere, and, for this
      purpose, our choice of $\chi(z)$ is as good as any other. The redshift of each object
      is then perturbed with a Gaussian photo-$z$ error with $\sigma_z=0.03\,(1+z)$, and
      the point distribution is projected in different redshift bins.
      
      The projected distributions of some of these catalogs for a redshift bin $0.5 < z < 0.6$
      are shown in figure \ref{fig:maps} (right panel). Figure \ref{fig:hth_beta} shows the value
      of $H_2(\theta)$ and its error calculated from these catalogs for a the same bin, together
      with the $\Lambda$CDM result from the lognormal catalogs. As is evident from this figure,
      when projected, these catalogs still retain their inhomogeneous nature, and can be clearly
      distinguished from an asymptotically homogeneous $\Lambda$CDM model for values of $D_2$ that
      are remarkably close to 3. For instance, only with the results drawn from the bin
      $0.5 < z < 0.6$ we would be able to set the limit $D_2\gtrsim 2.95$.
      
    \subsection{LTB models}
      \begin{figure}
        \centering
        \includegraphics[width=0.45\textwidth]{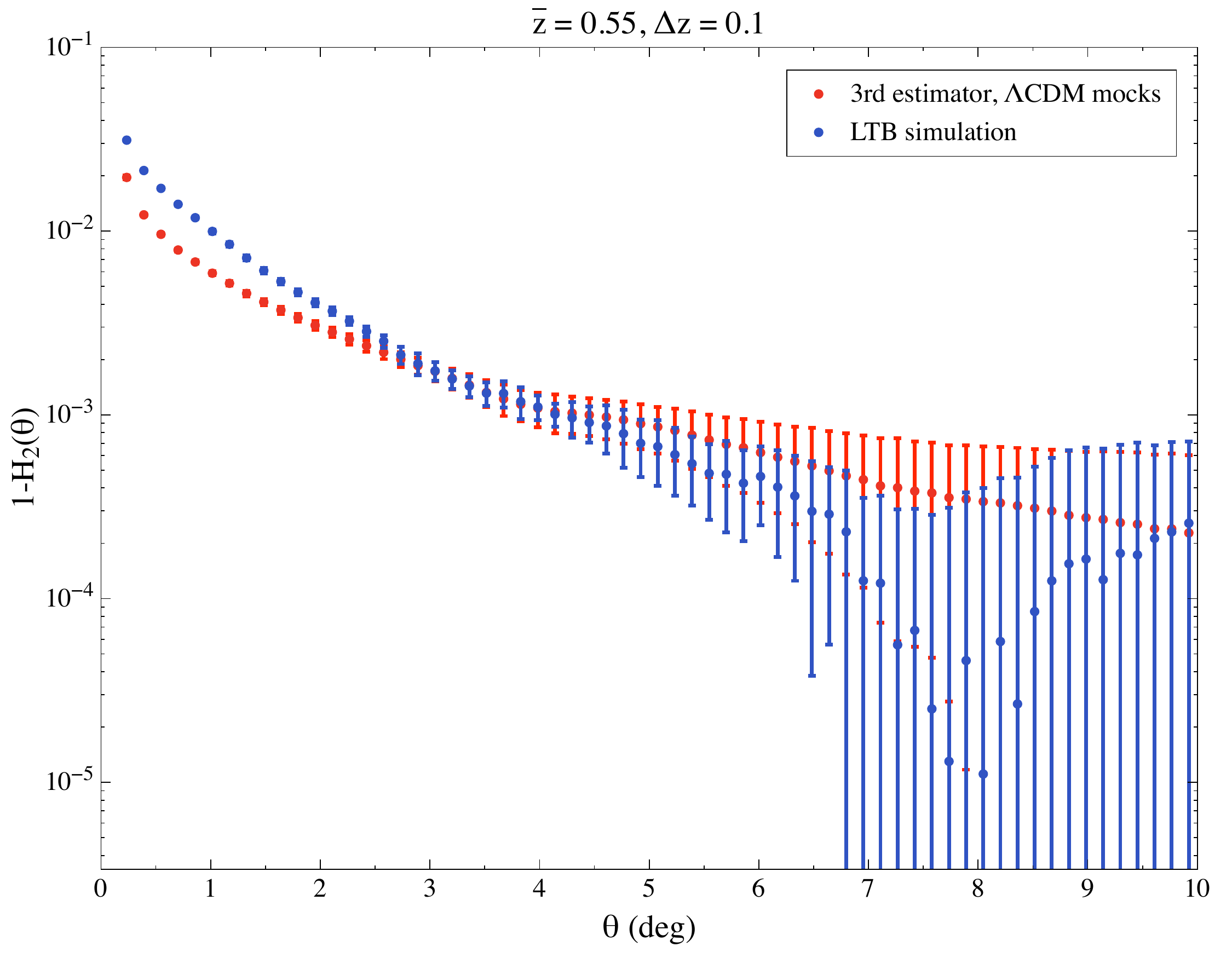}
        \caption{$\Delta H_2(\theta)$ as a function of $\theta$ for an LTB void model (blue)
                 compared to the $\Lambda$CDM value (red). Since the LTB metric preserves spherical
                 homogeneity around the central observer, the inhomogeneity of these models can not
                 be measured using the method described in this paper.}
        \label{fig:hth_ltb}
      \end{figure}
      In general there is no direct connection between the three dimensional fractal dimension
      $D_2$ and the homogeneity index $H_2$ of the projected data. An extreme example of this
      would be an inhomogeneous but spherically symmetric distribution in which the observer
      sits exactly at the centre of symmetry. Although the distribution is inhomogeneous in three
      dimensions ($D_2\neq3$), the central observer will measure a homogeneous distribution for
      the projected data ($H_2=1$).
      
      This is precisely the case of Lema\^itre-Tolman-Bondi (LTB) void models. A complete
      description of these models is out of the scope of this paper and the reader is
      referred to \citet{Clarkson:2012bg} for further information. For our purposes it is
      sufficient to know that, in an LTB model the observer is placed very close to the centre of a
      very large ($\mathcal{O}(1)\,{\rm Gpc}$) spherical underdensity. With this setup it is
      possible to
      reproduce many of the observational effects that can be adscribed to a Dark Energy component
      without introducing any exotic species or new physics in the model. The price to pay for this
      is relatively high, since in order to match the observed high isotropy of the distribution
      of CMB anisotropies, the observer is bound to be within a comparatively small distance
      ($\mathcal{O}(10)\,{\rm Mpc}$) from the centre of the void. LTB models have been tested
      against multiple cosmological observations \citep{GarciaBellido:2008nz,Bull:2011wi,
      Zumalacarregui:2012pq} and are basically ruled out. However, they provide an explicit example
      of an inhomogeneous model that can \emph{not} be distinguished from a homogeneous
      distribution with our method. This is shown in figure \ref{fig:hth_ltb}, in which we compare
      the homogeneity index $H_2(\theta)$ measured from the $\Lambda$CDM mock catalogs with the
      values measured from an N-body simulation of an LTB model. These simulations are described in
      \citet{Alonso:2010zv}, and the data shown in figure \ref{fig:hth_ltb} correspond to the
      best resolved simulation, labelled $\mathcal{H}$ in the aforementioned paper. The errors
      shown for the LTB data points have been calculated by splitting the simulation into 8 octants
      of the sky and then calculating the standard deviation of the 8 subsamples.
      
  \section{Summary \& Discussion}
  \label{sec:discussion}
    In this paper we have studied the possibility of measuring the transition to homogeneity using
    photometric redshift catalogs. The method presented here is an extension of the usual fractal
    studies that have previously been performed using three-dimensional distances by several
    collaborations. Photometric redshift uncertainties erase much of the clustering information
    along the radial direction. Thus, our method is based on measuring the fractality of the
    projected galaxy distribution, using only angular distances. This method is assumption-free,
    since it relies only on observable quantites (as opposed to three-dimensional distances, which
    require a fiducial cosmological model), and in this sense provides a way to test the
    Cosmological Principle in a model-independent way. In the era of precision cosmology, testing
    this fundamental assumption is extremely important, and the upcoming galaxy surveys, covering
    large volumes of the Universe, will make this possible.
    
    We have tested that our method is not biased by the use of random catalogs to correct for
    artificial effects induced on the observed galaxy distribution. We have done so by using our
    method on different synthetic inhomogeneous catalogs. We have verified that, not only is our
    method unbiased in practice, but it is in fact capable of discriminating some fractal models
    with relatively large fractal dimensions, in spite of the loss of information due to the
    radial projection. Our method is unable to detect the large-scale inhomogeneity along the line
    of sight, and therefore can not be used to constraint a particular type of ``malicious''
    inhomogeneous models preserving the isotropy around a central observer. 
    
    We have modelled and studied how different effects would affect the measurement of the angular
    homogeneity index $H_2(\theta)$ in a $\Lambda$CDM cosmology. We have studied the influence of
    the redshift bin width, photometric redshift errors, bias, non-linear clustering, and surveyed
    area. The level to which a given survey will be able to constrain the transition to homogeneity
    will depend mainly on two factors:
    \begin{itemize}
      \item The total surveyed area: this regulates the size of the statistical uncertainties.
      \item The compactness of the surveyed region: this determines the largest angular scale that
            can be measured.
    \end{itemize}
    In particular, a DES-like survey should be able to easily discriminate certain fractal models
    with fractal dimensions as large as $D_2=2.95$. We believe that this method will have relevant
    applications for upcoming large photometric redshift surveys, such as DES or LSST.

  \section*{Acknowledgements}
  \label{sec:acknowledgements}
  We acknowledge useful discussions with Luca Amendola, Pedro Ferreira, Valerio Marra and Guido W. 
  Pettinari. We thank the Spanish Ministry of Economy and Competitiveness (MINECO) for funding
  support through grants FPA2012-39684-C03-02 and FPA2012-39684-C03-03 and through the Consolider
  Ingenio-2010 program, under project CSD2007-00060. DA is supported by ERC grant 259505. ABB
  acknowledges support from the DFG project TRR33 ``The Dark Universe".

  \appendix
  \section{Lognormal mock galaxy catalogs}
  \label{sec:lognormals}
    For the analysis described in this paper we generated 100 lognormal realizations using the
    following method:
    \begin{enumerate}
      \item Consider a cubic box of side $L$ and divide it into $N_{\rm side}^3$
            cubical cells of size $l_c\equiv L/N_{\rm side}$. This will determine
            the scales available in the catalog: $2\pi/L \lesssim k \lesssim 2\pi/l_c$.
      \item We generate a realization of the Fourier-space Gaussian overdensity field at
            $z=0$ by producing Gaussian random numbers with variance
            \begin{equation}
              \sigma^2(k)\equiv\left(\frac{L}{2\pi}\right)^3P(k).
            \end{equation}
            This is done in a Fourier-space grid for ${\bf k}={\bf n}\,2\pi/L$ with
            $-N_{\rm side}/2 \leq n_i \leq N_{\rm side}/2$.
            
            At the same time, the $z=0$ velocity potential can be calculated from 
            the overdensity field as
            \begin{equation}
              \varphi_{\bf k}(z=0) = f_0\,H_0\,\frac{\delta_{\bf k}(z=0)}{k^2}
            \end{equation}
      \item Transform these fields to configuration space using a Fast Fourier Transform,
            and calculate the radial velocity at each cell by projecting the gradient
            of the velocity potential along the line of sight (LOS) (the direction of the
            LOS will depend on the position of the observer inside the box). This will yield
            the Gaussian overdensity $\delta_G$ and radial velocity $v_r$ fields at $z=0$. Note
            that, at this stage, we are assuming that the velocity field is purely irrotational,
            and that all vector contributions have died away.
      
      \item Calculate the overdensity field and radial velocity in the lightcone by computing
            the redshift to each cell through the distance-redshift relation
            \begin{equation}
              \chi(z)=\int_0^z\frac{dz'}{H(z')},
            \end{equation}
            and evolving the fields self-similarly to that redshift.
            
            At the same time we may perform the lognormal transformation on the Gaussian
            overdensity field. Thus, in a cell at ${\bf x}$ with redshift $z({\bf x})$, the
            overdensity and radial velocity are given by
            \begin{align}
              & 1+\delta({\bf x})=\exp\left[G(z)\delta_G({\bf x},z=0)-
                                            G^2(z)\,\sigma_G^2/2\right],\\
              & v_r({\bf x})=\frac{f(z)H(z)D(z)}{(1+z)\,f_0\,H_0}\,v_r({\bf x},z=0),
            \end{align}
            where $\sigma_G^2\equiv\langle\delta_G^2\rangle$ is the variance of the Gaussian
            overdensity at $z=0$ and the factor $G(z)\equiv D(z)\,b(z)$ accounts both for
            the growth of perturbations and a possible linear galaxy bias $b$.
            
      \item Calculate the mean number density of objects in each cell as $n({\bf x})\equiv
            \bar{n}(z)\,(1+\delta({\bf x}))$, where $\bar{n}(z)$ is the desired redshift
            distribution. The number of galaxies in each cell is calculated by generating
            a Poisson random number with mean $n({\bf x})$. These galaxies are placed at 
            random within each cell (thus losing any information about the clustering on
            scales below the cell resolution). The redshift of each galaxy is calculated using
            the distance-redshift relation and RSDs are produced by perturbing this redshift
            with $\Delta z_{\rm RSD} = (1+z)\,v_r({\bf x})$.
    \end{enumerate}

\end{document}